\documentclass[preprint,hidelinks,onefignum,onetabnum]{siamart251216}

\usepackage{amsfonts}
\usepackage{algorithm}
\usepackage{algpseudocode}
\usepackage{enumitem}
\usepackage{array}
\usepackage{multirow}

\algrenewcommand\algorithmicrequire{\textbf{Input:}}
\algrenewcommand\algorithmicensure{\textbf{Output:}}

\usepackage{hyperref}
\usepackage{graphicx}

\numberwithin{equation}{section}
\ifpdf
  \DeclareGraphicsExtensions{.eps,.pdf,.png,.jpg}
\else
  \DeclareGraphicsExtensions{.eps}
\fi

\newsiamremark{remark}{Remark}
\newsiamremark{hypothesis}{Hypothesis}
\crefname{hypothesis}{Hypothesis}{Hypotheses}
\newsiamthm{claim}{Claim}

\headers{Evaluating Cooling-Center Coverage using PH}{Erin O'Neil and Sarah Tymochko}

\title{Evaluating Cooling-Center Coverage Using Persistent Homology of a Filtered Witness Complex}

\author{Erin O'Neil\thanks{Program in Applied \& Computational Mathematics, Princeton University, Princeton, NJ 
(\email{erinoneil@princeton.edu}).}
\and Sarah Tymochko\thanks{Department of Mathematics and Computer Science, College of the Holy Cross, Worcester, MA
  (\email{stymochko@holycross.edu}).}
}

\usepackage{amsopn}

\makeatletter
\newcommand*{\addFileDependency}[1]{% argument=file name and extension
  \typeout{(#1)}% latexmk will find this if $recorder=0 (however, in that case, it will ignore #1 if it is a .aux or .pdf file etc and it exists! if it doesn't exist, it will appear in the list of dependents regardless)
  \@addtofilelist{#1}% if you want it to appear in \listfiles, not really necessary and latexmk doesn't use this
  \IfFileExists{#1}{}{\typeout{No file #1.}}% latexmk will find this message if #1 doesn't exist (yet)
}
\makeatother

\ifpdf
\hypersetup{
  pdftitle={Evaluating Cooling-Center Coverage Using Persistent Homology of a Filtered Witness Complex},
  pdfauthor={E. O'Neil and S. Tymochko}
}
\fi

\begin{document}

\maketitle

% REQUIRED
\begin{abstract}
In light of the increase in frequency of extreme heat events, there is a critical need to develop tools to identify geographic locations that are at risk of heat-related mortality. This paper aims to identify locations by assessing holes in cooling-center coverage using persistent homology (PH), a method from topological data analysis (TDA). Persistent homology has shown promising results in identifying holes in coverage of specific resources. We adapt these methods using a witness complex construction to study the coverage of cooling centers. We test our approach on four locations (central Boston, MA; central Austin, TX; Portland, OR; and Miami, FL) and use death times, a measurement of the size and scale of the gap in coverage, to identify most at risk regions. For comparison, we implement a standard technique for studying the risk of heat-related mortality called a heat vulnerability index (HVI). The HVI is a numerical score calculated for a geographic area based on demographic information.  PH and the HVI identify different locations as vulnerable, thus indicating a potential value of assessing vulnerability from multiple perspectives. By using the regions identified by both persistent homology and the HVI, we provide a more holistic understanding of coverage.
\end{abstract}

% REQUIRED

% REQUIRED

\section{Introduction}

The Environmental Protection Agency has identified extreme heat as the modern leading cause of weather-related mortality in the United States \cite{EPA_2023a}. The problem of heat-related mortality is expected to remain a pressing concern, especially if proactive measures are not adopted. Meehl and Tebaldi’s global coupled climate model \cite{Meehl} showed that heat waves in North America are projected to intensify, occur more frequently, and persist longer in the latter half of the 21st century compared to the latter half of the 20th century. This escalation in heat intensity is linked to rising levels of greenhouse gases.
There is evidence that the warmest day of the year will increase by 4-6\textdegree C (7.2-10.8\textdegree F) by the end of the century \cite{Pierce}. Heat accumulation is expected to be greater in urban areas due to human activity and construction, a phenomenon coined ``the urban heat island effect" \cite{Yang}.

Exposure to extreme heat can result in dire health outcomes including heat stroke, the exacerbation of existing medical conditions, permanent neurological damage, and, in the worst cases, loss of life \cite{Bouchama}. 
Often, the term ``excess deaths" is used when reporting the quantity of heat-related fatalities as they are often preventable. For instance, a single heat-wave event in Europe resulted in approximately 22,000-45,000 excess deaths in 2003 \cite{Bouchama}.
There is a critical need to develop tools to help locate vulnerable populations and identify protective measures against these heat-related deaths.

In a study that aimed to identify such protective measures, Bouchama et al.\ found that visiting air conditioned environments was strongly associated with better health outcomes \cite{Bouchama}. 
In fact, the meta-analysis showed that those who visited air conditioned places had an approximately 66\% lower likelihood of heat-related mortality. 
However, approximately 28.6\% of housing units in the United States do not have air conditioning \cite{Kim}. 
Furthermore, certain populations are more likely to lack air conditioning due to limited financial and social resources \cite{Kim}. 
Therefore, it is important for cities to offer free, accessible, public cooling centers, such as libraries, community centers, and senior centers \cite{Kim} during extreme heat events. 
It is imperative not only to increase the quantity of cooling centers but also to strategically position them in areas of highest necessity, taking into account factors such as the demographics within the city and existing cooling center locations. 
Recent work shows that the distribution of cooling centers is not always optimized to maximize access \cite{Kim}.

\subsection{Related Work and Contributions} \label{ssec:contributions}

Our goal is to study the coverage of cooling centers throughout a city and identify any holes in coverage.
Currently, a variety of computational tools exist to evaluate a region's cooling center coverage. 
One technique to evaluate coverage is the catchment area method which involves selecting a cut-off distance and evaluating who is within that distance of a given cooling center \cite{Kim, Nayak}. 
However, this approach alone overlooks important demographics that may be helpful in determining areas of greatest risk. 
For instance, a neighborhood of primarily senior citizens with little tree canopy may need to be treated differently than other neighborhoods. 
To account for this, many studies compute a social vulnerability index for a given geographic area based on demographic variables that influence an individual's ability to overcome environmental hazards \cite{Cutter}. 
In the context of extreme heat risk, they are often called \textit{heat vulnerability indices} (HVIs). 

There are multiple methods to calculate an HVI for a given region; see \cite{HVI_review1, HVI_review2} for a review of these approaches. 
Some previous studies utilize a combination of HVI maps and geospatial statistics to inform a maximal coverage location problem for finding ``optimal'' locations for new cooling centers.
For example, work by Bradford et al.\ \cite{bradford_heat_2015} and Fraser et al.\ \cite{Fraser} use an optimization approach that is informed by the HVI data. 
These methods, however, lack interpretability as they use proprietary ArcGIS tools.
Further, they rely on choosing a fixed distance threshold like the catchment area method.

A tool from topological data analysis (TDA), persistent homology (PH), provides a complementary perspective to these existing techniques. 
Firstly, TDA can evaluate coverage across a range of parameter values which bypasses the arbitrary selection of a cut-off distance. 
Second, many traditional methods, such as those that rely solely on the information provided by an HVI, treat each census block as an isolated ``island'' when assessing the risk of its residents. 
In contrast, PH is able to incorporate spatial information by considering cooling centers in neighboring census blocks. 
Third, our methodology considers cooling centers not only within city limits, but also in neighboring cities.\footnote{See Appendix \ref{appendix:data cooling centers} for a detailed explanation of the data and how it was collected.} 
In reality, individuals who live on the boundary of a given city may reside closer to cooling centers in neighboring cities than in their own. 
We therefore allow for the assumption that in an extreme heat event, individuals may visit neighboring cities to find relief. 

Our method is inspired by that of Hickok et al.\ \cite{Voting} which uses TDA to infer holes in voting site coverage. 
However, we use a different construction of the mathematical representation of space.
Other researchers have also addressed questions of resource coverage using PH. 
In particular, Gonzalez-Cruz et al.\ \cite{cruz-gonzalez2025sexual_healthcare} used PH to quantify the accessibility of Planned Parenthood and other health care clinics in California to identify regions lacking adequate health care facilities or those vulnerable in the case of clinics being defunded.
Persistent homology has shown success in many applications to spatial data beyond problems in resource coverage \cite{Snowflakes,hickok2022analysis,Feng_Porter_2021,Corcoran_Jones_2023,de2007coverage}.

\subsection{Organization of this paper}

In Section~\ref{sec:background} we introduce the relevant background from TDA; specifically, we introduce persistent homology.
In Section~\ref{sec:methods}, we present our persistent homology-based method as well as the HVI we use for comparison. 
We present and analyze the results in Section~\ref{Results} and conclude with discussions of implications and limitations of our work in Section~\ref{sec:conclusions}.

\section{Background} \label{sec:background}

Here we present a brief overview of topological data analysis and, in particular, persistent homology.
We first begin with some preliminary definitions in Section~\ref{ssec:defs}, followed by an introduction to persistent homology in Section~\ref{ssec:PH}.
We point the interested reader to \cite{Dey_Wang_2022} for a more formal treatment of this material.

\subsection{Definitions} \label{ssec:defs}

A simplicial complex is a shape constructed from lower dimensional building blocks such as vertices ($0$-simplices), edges ($1$-simplices), triangles ($2$-simplices), and higher dimensional analogs.
In general, a $k$-\textit{simplex} $\sigma$ is the convex hull of $k+1$ linearly independent points.
A \textit{face} $\tau$ of a simplex $\sigma$ is the convex hull of a subset of the points in $\sigma$; if $\tau$ is a face of $\sigma$, we write $\tau \subseteq \sigma$.
A $k$-simplex $\sigma$ can be identified uniquely by the set of $0$-simplices $\tau$ such that $\tau \subseteq \sigma$.
A \textit{simplicial complex} $\mathcal{K}$ is a set of simplices that satisfy two conditions: (1) $\mathcal{K}$ is closed under taking faces (i.e., if $\sigma \in \mathcal{K}$ and $\tau\subseteq \sigma$, then $\tau \in \mathcal{K}$) and (2) $\mathcal{K}$ is closed under intersections (i.e., if $\sigma_1, \sigma_2 \in \mathcal{K}$, then $\sigma_1 \cap \sigma_2 \in \mathcal{K}$).
A \textit{filtered simplicial complex} is a simplicial complex $\mathcal{K}$ and a function $f:\mathcal{K}\to \mathbb{R}$ such that $f(\tau)\leq f(\sigma)$ for $\tau \subseteq \sigma$.
From a filtered simplicial complex, one can build a nested sequence of simplicial complexes
\begin{equation} \label{eqn:filtration}
    \mathcal{K}_{\alpha_0} \subseteq \mathcal{K}_{\alpha_1} \subseteq \cdots \subseteq \mathcal{K}_{\alpha_n} \,,
\end{equation}
where $\alpha_0 < \alpha_1 < \cdots < \alpha_N$ and
$\mathcal{K}_n = \{ \sigma \in \mathcal{K} : f(\sigma) < \alpha_n \}$.
Equation~\ref{eqn:filtration} defines a \textit{filtration}.
We call $\alpha$ the \textit{filtration parameter}.

\subsubsection{Constructing a Simplicial Complex from Data}

The first step in computing the ``shape'' of data is to first represent the data as a (filtered) simplicial complex.
This can be done in numerous ways, two of the most common being the \v{C}ech or Vietoris--Rips complexes \cite{Dey_Wang_2022}. 
Given a collection of points $X=\{x_1, \ldots, x_n\}$ in a metric space $(M,d)$, the \textit{\v{C}ech complex} at filtration parameter $\alpha$ is the simplicial complex with vertices $X$ and simplices $\{x_{i_0}, \ldots x_{i_k}\}$ if $\bigcap_j B(x_{i_j}, \alpha)$ is nonempty, where $B(x, \alpha) = \{y \in M : d(x,y)\leq \alpha\}$. 
The \v{C}ech complex is topologically equivalent to the nerve of the union of balls, $\bigcup_j B(x_{i_j}, \alpha)$ by the Nerve theorem \cite{borsuk1948imbedding} so long as the balls are convex.
The \v{C}ech complex can be used to create a filtration by using an increasing sequence of filtration parameter values, $\alpha_0 < \alpha_1 <\ldots < \alpha_N$.

The \v{C}ech complex is very useful in theory, but in practice is very computationally prohibitive. 
It is common to instead use the \textit{Vietoris--Rips (VR) complex} as it is an approximation of the \v{C}ech complex and is notably faster to compute.\footnote{Computing the \v{C}ech complex requires knowledge of the metric space in order to check if the intersection of $k$ balls is nonempty. 
In comparison, the VR complex requires only the pairwise distances between all points.}
The VR complex at filtration parameter $\alpha$ is the simplicial complex with vertices $X$ and simplices $\{x_{i_0}, \ldots x_{i_k}\}$ if $d(x_{i_j},x_{i_\ell}) < 2\alpha$ for all $j$ and $\ell$.
Just as with the \v{C}ech complex, the VR complex gives rise to a filtration when computed for an increasing sequence of filtration parameter values.

Here we use a different approach for creating a simplicial complex from data. In particular, we use the \textit{witness complex} $\mathcal{W}$ because it allows for the incorporation of two sets of points, the landmarks $L$ and the witnesses $W$ \cite{desilvacarlsson2004}.
Often the landmarks are chosen as a subset of the witnesses, however, this is not a requirement.\footnote{The VR complex remains computationally prohibitive for large point sets. The witness complex can capture similar topological structure in a way that is more computationally tractable. In this setting, the full point cloud serves as the set of witnesses while a subset of the points serve as the landmarks. 
The resulting complex has fewer simplices than a Vietoris--Rips complex, thus speeding up computations.
This, however, is not our motivation for using the witness complex.} 
In our case, the landmarks and witnesses will come from two separate sets of points; the landmarks represent neighborhoods throughout a city and the witnesses represent the locations of the cooling centers. 
The landmarks form the vertices of the simplicial complex and witnesses are used to determine which higher dimensional simplices are present. 
That is, a $k$-simplex for $k>0$ is in the complex if it is ``witnessed'' by a point in $W$.  
This ``witness'' relation can be defined in different ways to create variants such as a weak or strong witness complex \cite{desilvacarlsson2004, Alexander_Bradley_Meiss_Sanderson_2015, sanderson2018topological}.
We define a \textit{filtered witness complex} as a witness complex paired with a filter function $f:\mathcal{W}\to \mathbb{R}$ where $f(\sigma)$ is the distance from $\sigma$ to the closest witness.
This gives rise to a filtration using the same definition as in Eqn.~\ref{eqn:filtration}.
We define our construction of the witness complex in Section \ref{Witness Complex}.

\subsection{Persistent Homology} \label{ssec:PH}

\textit{Homology} is an invariant from algebraic topology used to quantify holes in different dimensions. 
A 0-dimensional (0D) hole represents a connected component and a 1-dimensional (1D) hole represents a cycle or a loop. 
Given a simplicial complex, one can compute homology; given a filtration, one can compute \textit{persistent homology}.
At each step in the filtration, connected components may appear and merge, and holes may form and fill in.
A component or a hole is said to be \textit{born} in the first step of the filtration in which it appears. 
A component is said to have \textit{died} when it merges with another connected component, while a hole dies when it is filled in. 
More formally, the \textit{birth time} of a homology class is $\alpha_b$ if the homology class appears in $\mathcal{K}_{\alpha_b}$ but not $\mathcal{K}_{\alpha_i}$ for $i<b$. 
Similarly, the \textit{death time} of a homology class is $\alpha_d$ for $d>b$ if the homology class is present in $\mathcal{K}_{\alpha_{d-1}}$ but not in $\mathcal{K}_{\alpha_{d}}$.
When two components merge, the older (i.e. the one born earlier) persists while the younger dies; this is known as the elder rule \cite{Edelsbrunner_Harer_2009}.

The \textit{birth simplex} is the first simplex to appear in the filtration that creates a homology class, while the \textit{death simplex} is the last simplex added that kills the homology class.
In general, a death simplex may not be unique; for example, if three vertices are all born at time 0 and the three pairwise edges are all added at the same filtration value, all three edges could be considered the death simplex. 
In real datasets, this is rarely an issue: distances are typically distinct, so edges and higher-dimensional simplices enter at unique filtration values.
Another concern with death simplices is that they are not stable; that is, small perturbations in the data can cause the death simplex to change locations. 

Persistent homology results in a \textit{persistence diagram}, a multiset of points where a homology class that is born at $b$ and dies at $d$ is represented by the point $(b,d)$.
Since death times are always greater than birth times, all points in the persistence diagram fall above the line $b=d$.

\begin{figure}
\centering 
\includegraphics[scale = 0.45]{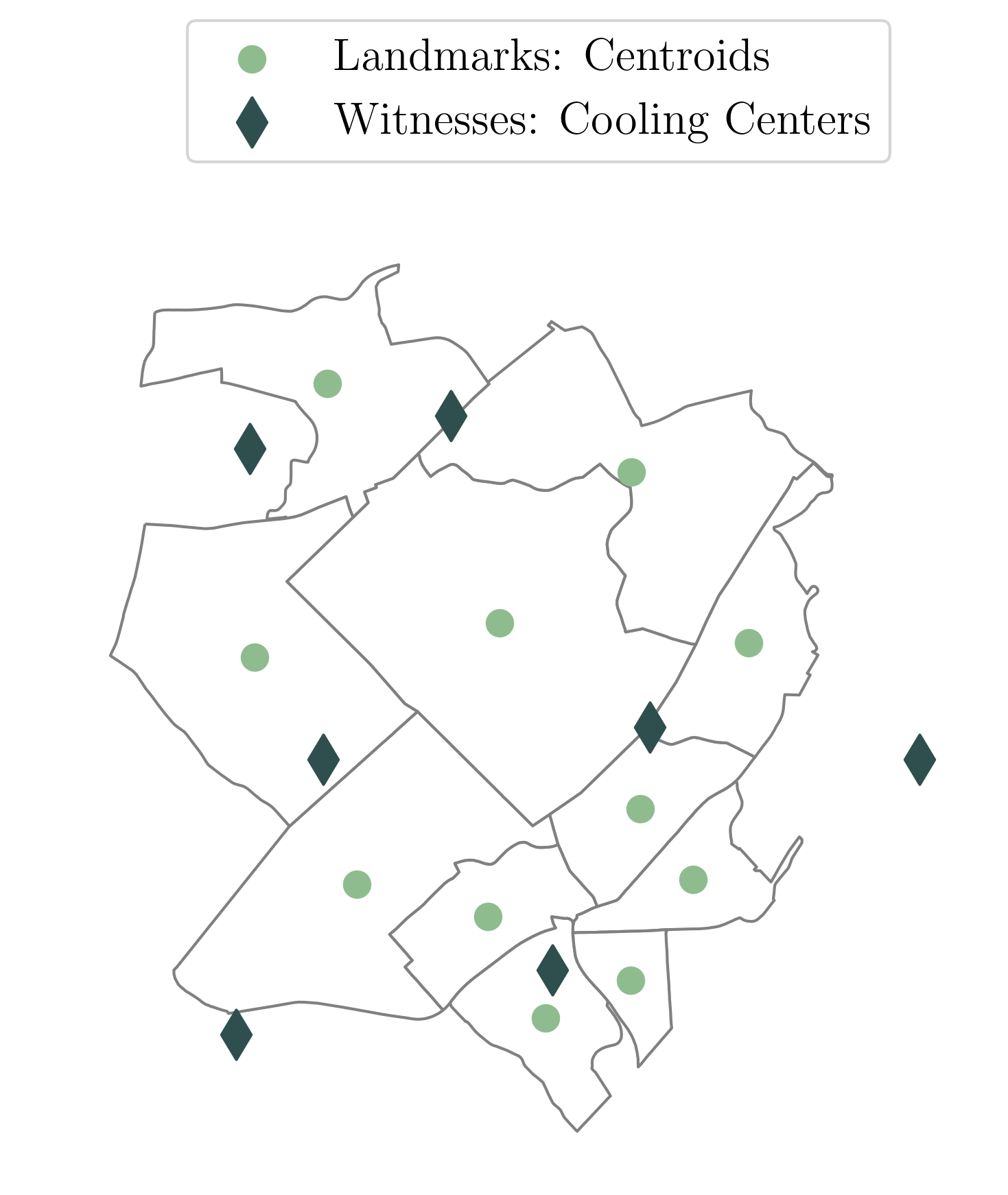}
\caption{An example of 11 census tract centroids (landmarks) and 7 cooling centers (witnesses). Note: this image was created based on a small subset of Boston, MA census tracts and the witness locations were generated randomly. To see the full city of Boston with the true locations of cooling centers, see Figure~\ref{data}.} \label{RandomExample}
\end{figure}

\section{Methods} \label{sec:methods}
In this section we present two methodologies for evaluating cooling center coverage. 
Section \ref{Witness Complex} introduces our topological approach using persistent homology of a filtered witness complex to identify the vulnerable locations. 
Section \ref{Score} defines a \textit{heat vulnerability index} (HVI), inspired by existing literature, to serve as a comparison to the topological approach. 

In order to represent the spatial layout of the city, we use geographic units that are used by the U.S. Census Bureau to aggregate data.
Our topological approach uses census blocks, the smallest geographic unit used by the Census Bureau, while the HVI was computed at the census tract level. 
This is because the demographic data needed to compute the HVI is only available at the census tract level. 
% In our topological
See Appendix \ref{AppendixData} for more details.

\subsection{Our Construction of the Witness Complex} \label{Witness Complex}

The witness complex is a simplicial complex constructed based on two sets of points: landmarks, $L = \{ \ell_i \}$, and witnesses, $W = \{w_j\}$.
We will construct it as follows:
\begin{definition*}
Let $L$ be the set of landmark points, $W$ be the set of witness points, and $\alpha\in \mathbb{R}^{\geq 0}$ be a scale parameter. We define the witness complex constructed from $L$ and $W$ at scale $\alpha$ as follows:
\begin{align*} 
    \mathcal{W}(L,W)_\alpha = \{\sigma : \text{there exists some $w\in W$ such that } d(l, w) \leq \alpha \text{ for all }\ell \in \sigma \} \cup L
\end{align*}
where $d(\ell, w)$ is the distance between landmark $\ell$ and witness $w$.
Note that equation above includes all landmark points as 0-simplices for all scales $\alpha\geq 0$. 
\end{definition*}

Intuitively, this is a simplicial complex constructed with landmarks as the vertices (all of which are born at $\alpha=0$), and simplices built from the sets of vertices contained within distance $\alpha$ from the same witness. 
Given an increasing sequence of $\alpha$ values 
\[
0< \alpha_0 < \alpha_1 < \alpha_2 < \cdots < \alpha_n
\]
then the sequence of simplicial complexes
\[
\mathcal{W}(L,W)_{0} \subseteq \mathcal{W}(L,W)_{\alpha_0} \subseteq \mathcal{W}(L,W)_{\alpha_1}  \subseteq \cdots \subseteq \mathcal{W}(L,W)_{\alpha_n}
\]
defines a filtration. 

\begin{figure}
\centering 
\includegraphics[width = 0.7\textwidth]{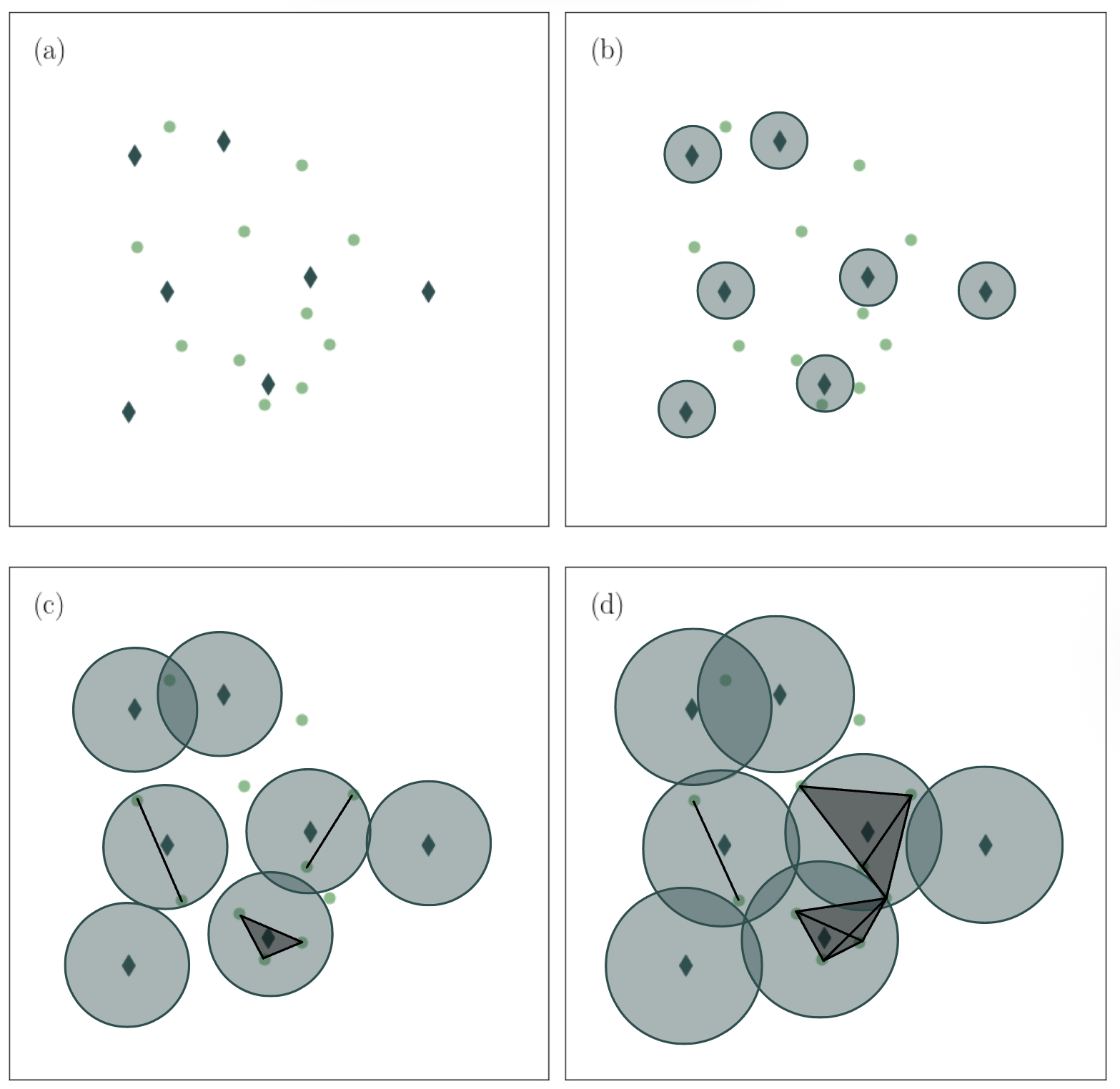}
\caption{Example of select steps in the filtration. As shown in Figure \ref{RandomExample}, light green points represent the centroids (landmarks) of the geographic region and dark green diamonds (witnesses) represent the cooling center locations. All vertices (landmarks) appear in (a). (b)-(d) show the simplicial complex (in black) at various stages of the filtration. Edges are drawn between vertices when they are within a distance $\alpha$ of the same witness. Green discs show the radii at that filtration value. 
Note that the witnesses are drawn for easier visualization but they are not vertices in the simplicial complex.
\label{Filtration}}
\end{figure}

In this work, the centroids of census blocks are used as the landmarks, and the locations of cooling centers are used as the witnesses. 
See Figure \ref{RandomExample} for an example.
This means, in the filtered witness complex, an edge is added between two landmarks when the corresponding census blocks are both within distance $\alpha$ of the same cooling center.
See Figure \ref{Filtration} as an example of how components evolve over select steps in a filtration.
When two landmarks are connected by an edge, in other words, when they have access to the same cooling center, we interpret this as the two census blocks having ``equal coverage.''

\subsubsection{Interpretation and Visualization} \label{ssec:deathsimps}

In order to further identify and visualize regions of poor coverage from our topological approach, we use the death simplices as defined in Section~\ref{ssec:PH}.\footnote{Another approach to identifying ``holes'' in coverage from 1-dimensional homology is using cycle representatives. However, because cycle representatives are not unique and finding ``optimal'' cycles is an open and ongoing area of research, we leave this as potential future work.}
Death simplices for $0$-dimensional homology are edges (1-simplices) which represent the edge that is added to connect two previously distinct components.
These edges connect components that previously had differing levels of cooling center coverage.
For $1$-dimensional homology, death simplices are triangles (2-simplices) which represent the final part of the hole to be covered.
These regions are the convex hull of three census blocks between which there is a lack of cooling centers.
A death simplex with a higher death time indicates worse coverage in that region.

As discussed previously, death simplices need not be unique in theory. 
However in this case, because simplices of dimension 1 or greater are added based on distances calculated between geographical points, then we can guarantee that at most one simplex is added at any given filtration value and thus the death simplices are unique.\footnote{When identifying the 0-dimensional death simplices, we recompute the persistence points after slightly perturbing the birth times of the $0$-simplices. 
This ensures that each 0-dimensional feature can be uniquely matched to the corresponding death simplex. 
The perturbations are carefully chosen so as not to impact the order in which simplices of dimension 1 or greater are added, thus not changing the death times of persistence points nor the corresponding death simplices.}

Details on the data we use can be found in Appendix~\ref{appendix:data centroids} and \ref{appendix:data cooling centers}. 
Our code for constructing the witness complex can be found on GitHub.\footnote{\href{https://github.com/erinponeil/Cooling_Center_Converage_TDA.git}{https://github.com/erinponeil/Cooling\_Center\_Converage\_TDA.git}}
We use the \texttt{GUDHI} library \cite{gudhi} for computing persistent homology.

\subsubsection{Distance Computation} \label{ssec:distance_computation}

We consider the fact that each city is in a different region of the country. 
This is an important consideration because distances, like a Euclidean distances, can quickly become inaccurate due to the failure to account for the curvature of the Earth. 
In other words, a projected map of the US may distort distances in some cities more than others.
To avoid the distortion of projected maps, we compute distances between latitude and longitude points using a \textit{straight-line} measurement along Earth's surface which is often referred to as the \textit{geodesic distance}. 
We believe this choice to be justifiable because some studies have shown straight-line measurements are an acceptable proxy to travel distances in urban areas \cite{Bliss, Boscoe, Jones}. 
Furthermore, in non-emergency situations, Boscoe et al.\ concluded that the difference between straight-line and driving estimates were inconsequential and, because of this, straight-line estimates are often used \cite{Boscoe, Nayak}. 
This measure of distance is then converted into units of kilometers for easier interpretation.

\begin{figure}
\centering 
\includegraphics[width = 0.98\textwidth]{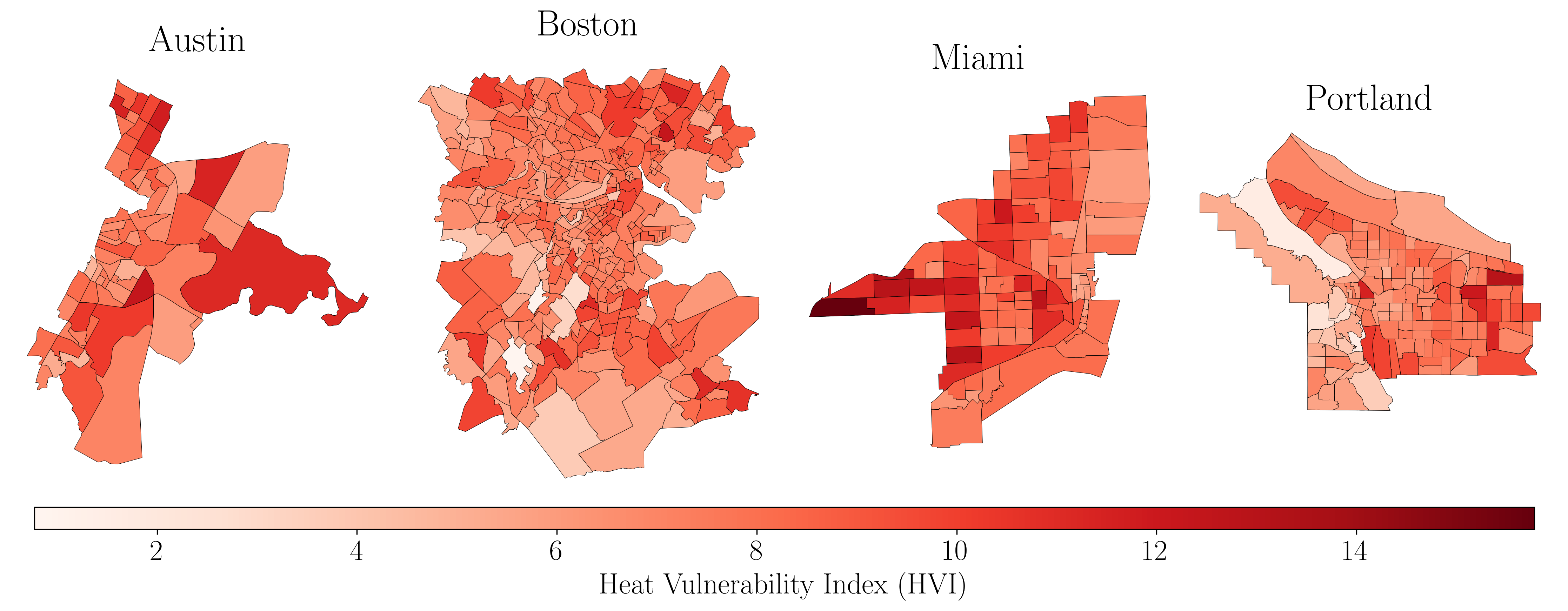}
\caption{Heat vulnerability index (HVI) score maps for the four regions of interest. 
} \label{fig:HVI_Score}
\end{figure}

\subsection{Heat Vulnerability Index (HVI)} \label{Score}

In order to compare our topological approach against standard methods for measuring heat vulnerability, we construct a \textit{Heat Vulnerability Index (HVI)} based on the existing literature.
Multiple methods exist for calculating HVIs in urban areas, including principal components analysis (PCA), ordered weighted averaging (OWA), hierarchical agglomerative clustering, theoretical component (TC), and unweighted $z$-scores, among others \cite{HVI_review1, HVI_review2}. 
For simplicity, we opt to use an unweighted $z$-score that is modeled after several other HVIs \cite{Christenson, Conlon, Fard}. 
In this study, the HVI map serves as a contextual overlay and comparison for our TDA analysis, providing insight into the demographics of the city under investigation.\footnote{One can construct their own contextual HVI maps using alternative variables, methods, and weights using the code in \href{https://github.com/erinponeil/Cooling_Center_Converage_TDA.git}{https://github.com/erinponeil/Cooling\_Center\_Converage\_TDA.git}.}

The HVI is based on characteristics of the demographics and environment that have been shown to exacerbate the likelihood that a person will suffer negative health consequences from an extreme heat event. 
Since the demographic data is not available at the census block level, we define this score on census tracts.
The characteristics of a census tract $T$ that we use are the typical afternoon temperature $F(T)$ in Fahrenheit,\footnote{Note that afternoon temperature data was collected at 3pm in the respective city using a combination of Sentinel-2 satellite data and sensors mounted on the bikes and cars of volunteers \cite{heatUSUrban}} the percentage of the area not covered by tree canopy $C(T)$, and the number of younger and older residents, $Y(T)$ and $S(T)$ respectively.\footnote{Note that the National Integrated Heat Health Information System defines ``younger'' to mean under 5 years old and ``older'' to mean over 65 years old \cite{heatUSUrban}.}
Because these variables have different units of measurement, we standardize them before creating a composite score. 
The HVI of census tract $T$ is
\begin{equation} \label{eqn:vulnerabilityscore}
HVI(T) = \frac{F(T)-\mu_F}{\sigma_F} + \frac{C(T)-\mu_C}{\sigma_C} + \frac{Y(T)-\mu_Y}{\sigma_Y} + \frac{S(T)-\mu_S}{\sigma_S} \,,
\end{equation}
where $\mu_F$ is the mean and $\sigma_F$ is the standard deviation of temperatures (other variables are defined analogously) across all census tracts in all cities studied in \cite{heatUSUrban}.
Because the standardization is done across all cities, we can compare the HVIs across cities.

Various factors influence an individual's risk of heat-related mortality. 
We select four variables that span demographic, geographic, and topographic features of the area. 
See Appendix~\ref{appendix:data score} for additional discussion of the HVI variables.
In particular, Figure~\ref{fig:HVI_figs_all} shows maps of each of the four variables individually before they are combined to make the HVI shown in Figure~\ref{fig:HVI_Score}.

\subsubsection{Evaluating Multicollinearity} \label{sssec:multicol}

Cutter et al.\ explain that an essential component of building a HVI involves testing for multicollinearity among demographic variables \cite{Cutter}. 
We examine the four variables used in this study and test for independence using variance inflation factor (VIF) \cite{multivar_data_analysis}.
The full VIF results can be found in the Appendix in Table \ref{table:VIF}, however we briefly discuss them here.
The VIF analysis for Austin and Miami indicate that the low amount of multicollinearity is acceptable for the proposed analysis for all four variables.
The results for Portland indicate the afternoon temperature and tree canopy have non-negligible multicollinearity and thus are likely not independent of each other.
Because of this, the HVI results for Portland should be considered with caution. 
We further analyze the independence and multicollinearity using pairwise correlations in Appendix~\ref{appendix:additional_data_analysis} and Figures~\ref{fig:HVI_histograms} and \ref{fig:HVI_correlationmatrices}.
This analysis reveals a strong pairwise correlation between the afternoon temperature and tree canopy in Boston. 
Thus, the HVI results for Boston must also be considered with caution.

\begin{figure}[t!]
\centering 
\includegraphics[width = 0.8\textwidth]{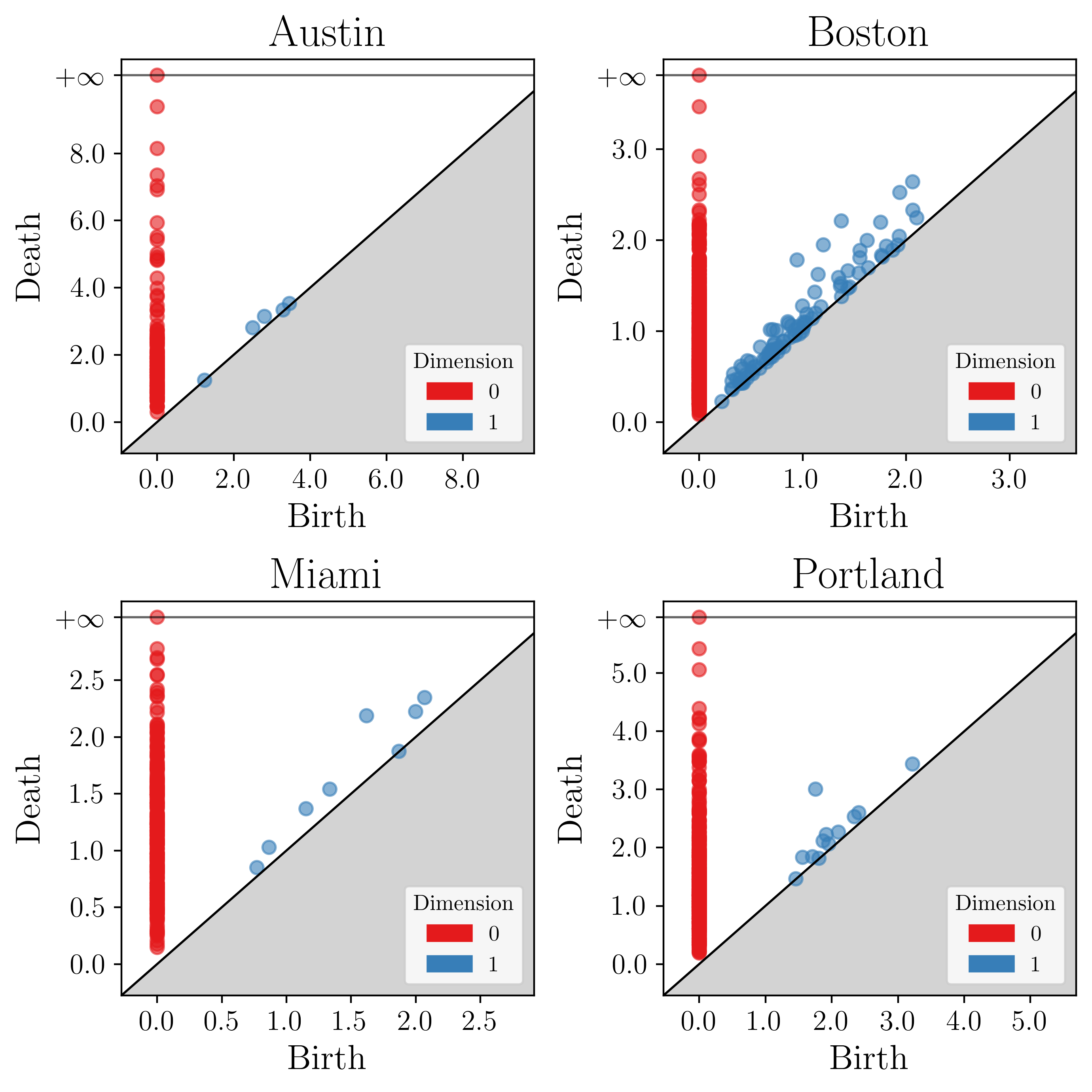}
\caption{Persistence diagram for our four geographic locations. The birth and death times are in units of kilometers. 
Note that the axes are not on the same scale.} \label{PD}
\end{figure}

\subsubsection{Limitations}

While the purpose of the HVI is to serve as a comparison to our topological approach, we will briefly discuss its limitations.
We limit our HVI to only include four variables for the sake of simplicity, however, the U.S. Urban Heat Island Mapping Campaign dataset contains several other variables that could be incorporated.
This includes features such as average morning and evening temperatures, the number of individuals living in poverty, the median income, and the percentage of the population that is a minority\footnote{Defined by the U.S. Urban Heat Island Mapping Campaign to mean non-white.}.
Because we chose two demographic variables, it is possible the HVI results are biased towards places with a higher population.
For example, places that are cooler but have a higher population of individuals $<5$ years old or $>65$ years old may have a higher HVI than a place that is warmer but has a lower population of vulnerable individuals.
Beyond which variables are included, there are modifications that could be made to the approach.
While the unweighted $z$-score approach is standard in the existing literature, it weighs each variable equally and it increases linearly with the increase of any one variable.
Because measuring heat vulnerability is multifaceted, it is unclear if these are valid assumptions.\footnote{One can explore the HVI and incorporate different weightings and combinations of variables using code in our Github repository. }

\begin{figure}[ht]
\centering 
\includegraphics[width=\textwidth]{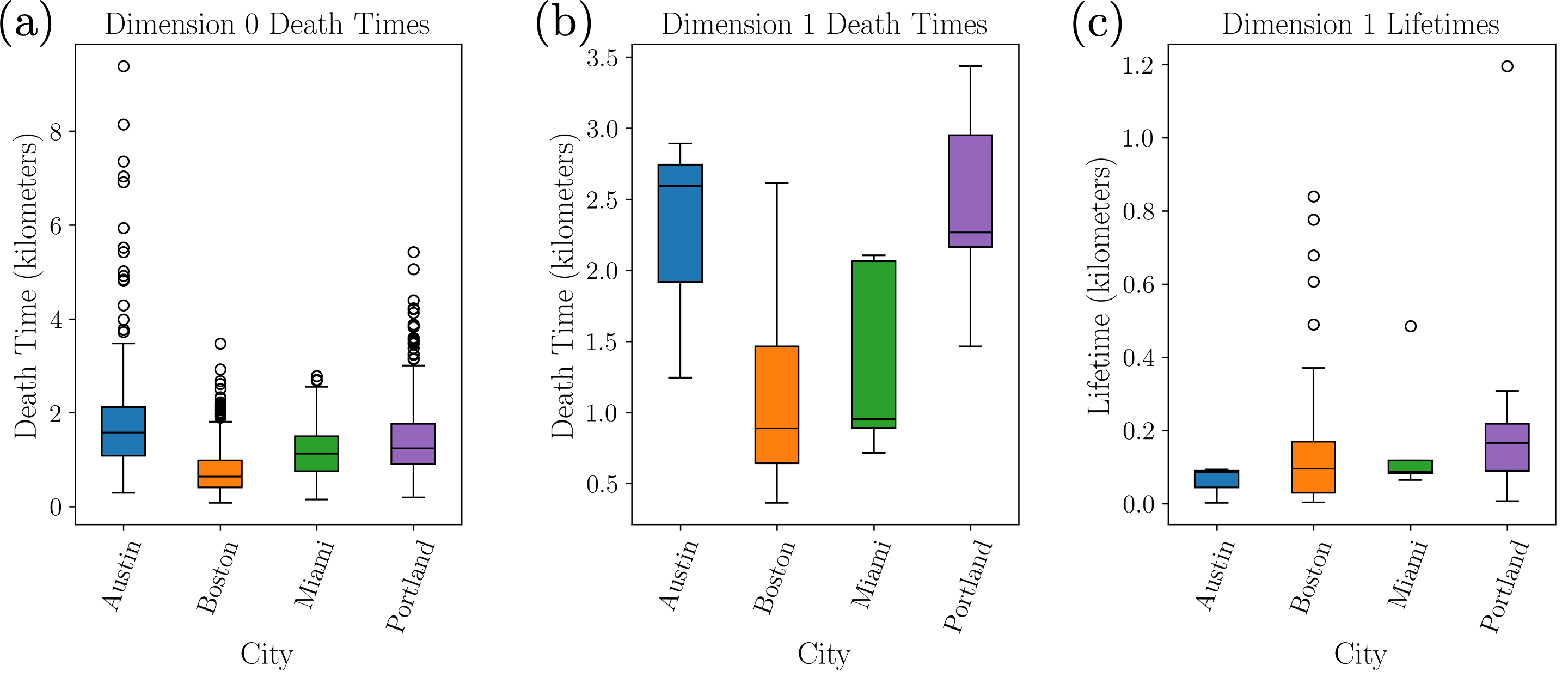}
\label{fig:boxplots}
\caption{Box plots for each of the four geographic locations of interest of (a) death times of the 0D homology classes, (b) death times of 1D homology classes, and (c) the lifetimes of the 1D homology classes.
A table of the corresponding means and standard deviations can be found in Table~\ref{tab:deathstats}.
}
\end{figure}

\section{Results}\label{Results}

In this section we will analyze the results of the HVI and our topological approach in Sections~\ref{ssec:hviresults} and \ref{ssec:toporesults} respectively, followed by a comparison of the two approaches in Section~\ref{ssec:compareresults}.

We test our methodology on four locations in the United States that offer unique climates and demographics. 
Those locations include central Boston, MA; central Austin, TX; Portland, OR; and Miami, FL. 
These cities differ in various ways, for example Miami and Boston are on the water, while Austin is landlocked. 
Additional differences include humidity, which can increase how warm a temperature feels, and the populations of vulnerable demographics such as older adults. 
All of the selected cities are vulnerable to the urban heat island effect  \cite{heatUSUrban}.

\subsection{HVI} \label{ssec:hviresults}

\begin{figure}[]
    \centering 
    \includegraphics[width=0.53\textwidth]{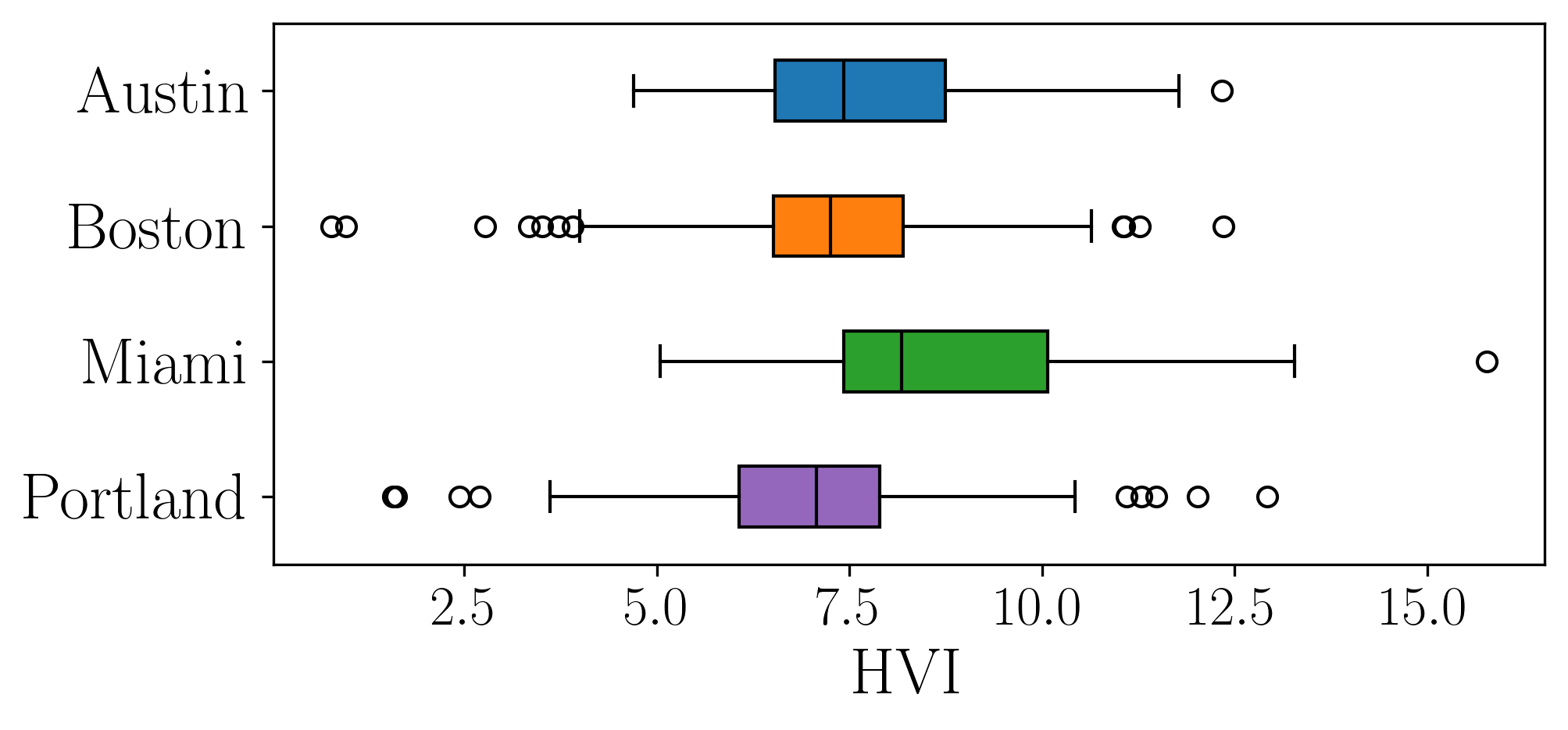} 
    \caption{Boxplot of the HVIs across all census tracts within a given city. A table of the corresponding means and standard deviations can be found in Table~\ref{tab:deathstats}.}
    \label{fig:hvibox}
\end{figure}

Based on the HVI maps in Figure~\ref{fig:HVI_Score}, we can identify the regions of higher HVI which indicate regions more vulnerable to heat related mortality. 
In particular, we can see individual census tracts in central Austin, northern Boston, western Miami, and eastern Portland with particularly high HVIs. 
We can see based on the box plots of the HVI values in Figure~\ref{fig:hvibox} that all cities have similar medians, although the HVI values in Miami are slightly higher. 
A table of statistics of the HVI values can be found in the last column of Table~\ref{tab:deathstats}. 
Additionally, Figure~\ref{fig:HVI_figs_all} shows heatmaps of each of the individual variables used in the HVI computation for each city. 
These can be used to identify regions where one particular indicator of vulnerability may be high even if the composite HVI is not. 

\subsection{Topological Approach} \label{ssec:toporesults}

\begin{table}[ht]
    \caption{Mean and standard deviation (SD) of: dimension 0 death times, dimension 1 death times, dimension 1 lifetimes, and HVI. Note that the units of death times and lifetimes are kilometers. The HVI is unitless due to it being an unweighted z-score.}
    \centering
    \begin{tabular}{l c  c  c  c}
        \hline
        \hline
         & \multicolumn{1}{c}{\textbf{0D Death}} & \multicolumn{1}{c}{\textbf{1D Death}} & \multirow{2}{6em}{\textbf{1D Lifetime}} & \multirow{2}{6em}{\centering \textbf{HVI}} \\
         & \textbf{Time} & \textbf{Times} & & \\
        % \cline{2-7}
        \textbf{City} & Mean $\pm$ SD & Mean $\pm$ SD & Mean $\pm$ SD & Mean $\pm$ SD \\
        \hline
        \hline
        Austin & 1.939 $\pm$ 1.428 & 2.812 $\pm$ 0.820 & 0.152 $\pm$ 0.141 & 7.819 $\pm$ 1.746 \\ 
        % \hline
        Boston & 0.745 $\pm$ 0.454 & 1.111 $\pm$ 0.567 & 0.146 $\pm$ 0.174 & 1.326 $\pm$ 1.499 \\
        % \hline
        Miami & 1.149 $\pm$ 0.519 & 1.677 $\pm$ 0.532 & 0.217 $\pm$ 0.156 & 8.715 $\pm$ 2.051 \\
        % \hline
        Portland & 1.418 $\pm$ 0.796 & 2.268 $\pm$ 0.530 & 0.259 $\pm$ 0.312 & 7.046 $\pm$ 1.805 \\
        \hline
    \end{tabular}
    \label{tab:deathstats}
\end{table}

We compute the PH of the filtered witness complex for each of the four locations of interest. Their persistence diagrams (PDs) are shown in Figure \ref{PD}. 
Notably, the PD for Boston shows more 1D homology class points than any other location. 
Figure \ref{fig:boxplots} plots the distributions of 0D and 1D death values for the four locations of interest.
The average death times (which are listed in Table~\ref{tab:deathstats}) are fairly similar across cities in both dimension 0 and 1; the average death time in dimension 0 is between 0.5 and 2 while the average in dimension 1 is between 1 and 2.5. 
It is worth noting that Austin has many more outliers with high 0-dimensional death time than any other city, possibly indicating more regions with greater risk.

For dimension 0, all features have a birth time of 0 so the death time is also the lifetime.
For dimension 1, however, the birth time is nonzero so we also compute the lifetime of the persistence points.
The birth time alone is not informative in the context of this application, however if the lifetime is small, it indicates that the hole fills in soon after it is formed so it may not be a notable hole in coverage.
Even though in dimension 1 Austin has the second highest average death time, the average lifetime is the smallest across all four cities, likely indicating the regions indicated by the 1-dimensional death simplices are less vulnerable in Austin than in other cities. 
On the other hand, Boston has the largest death times and lifetimes of 1-dimensional features across all four cities, likely indicating there are more notable holes in coverage in Boston than in Austin.

\begin{figure}
\centering 
\includegraphics[width=0.48\textwidth]{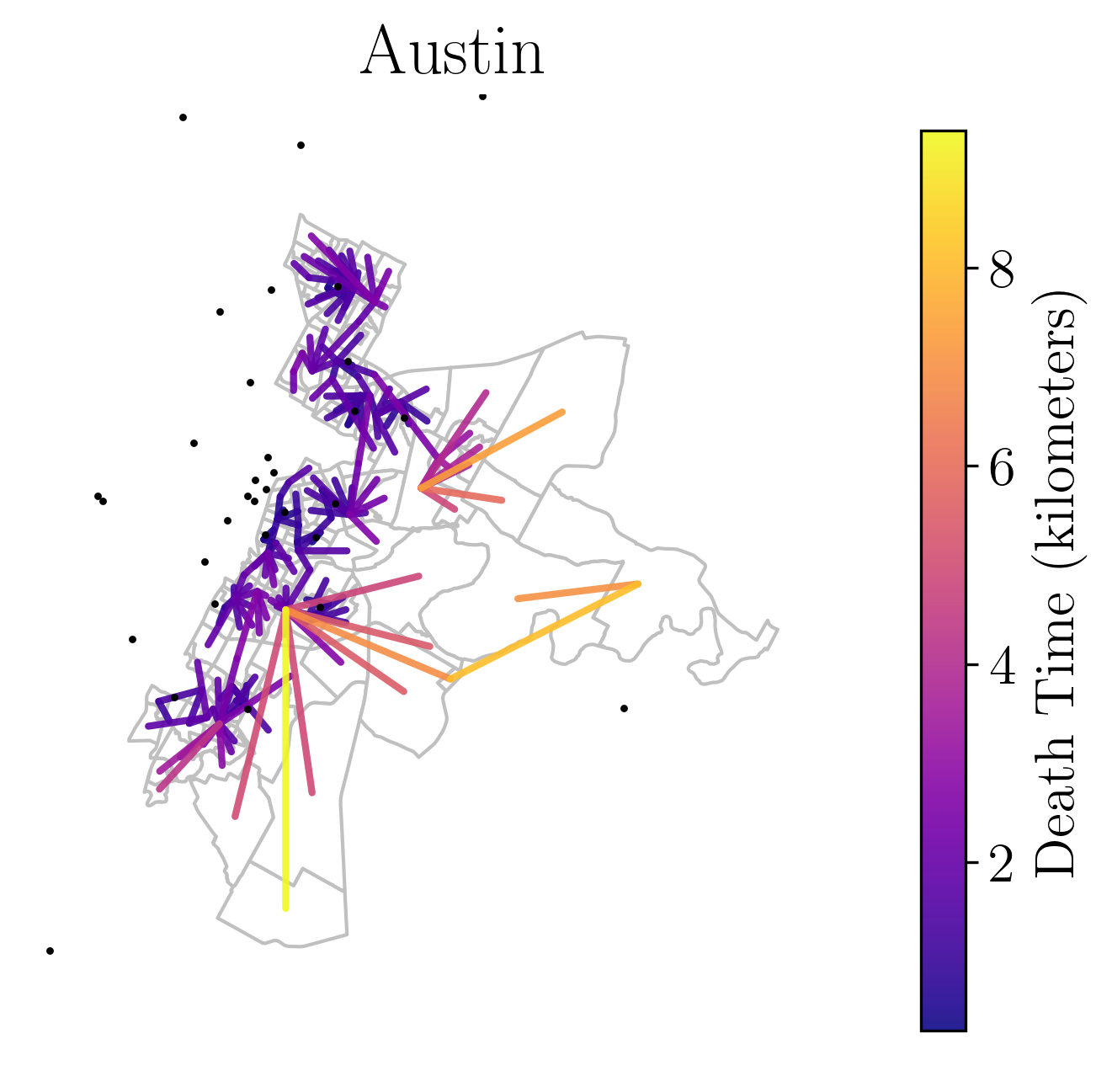}
\includegraphics[width=0.46\textwidth]{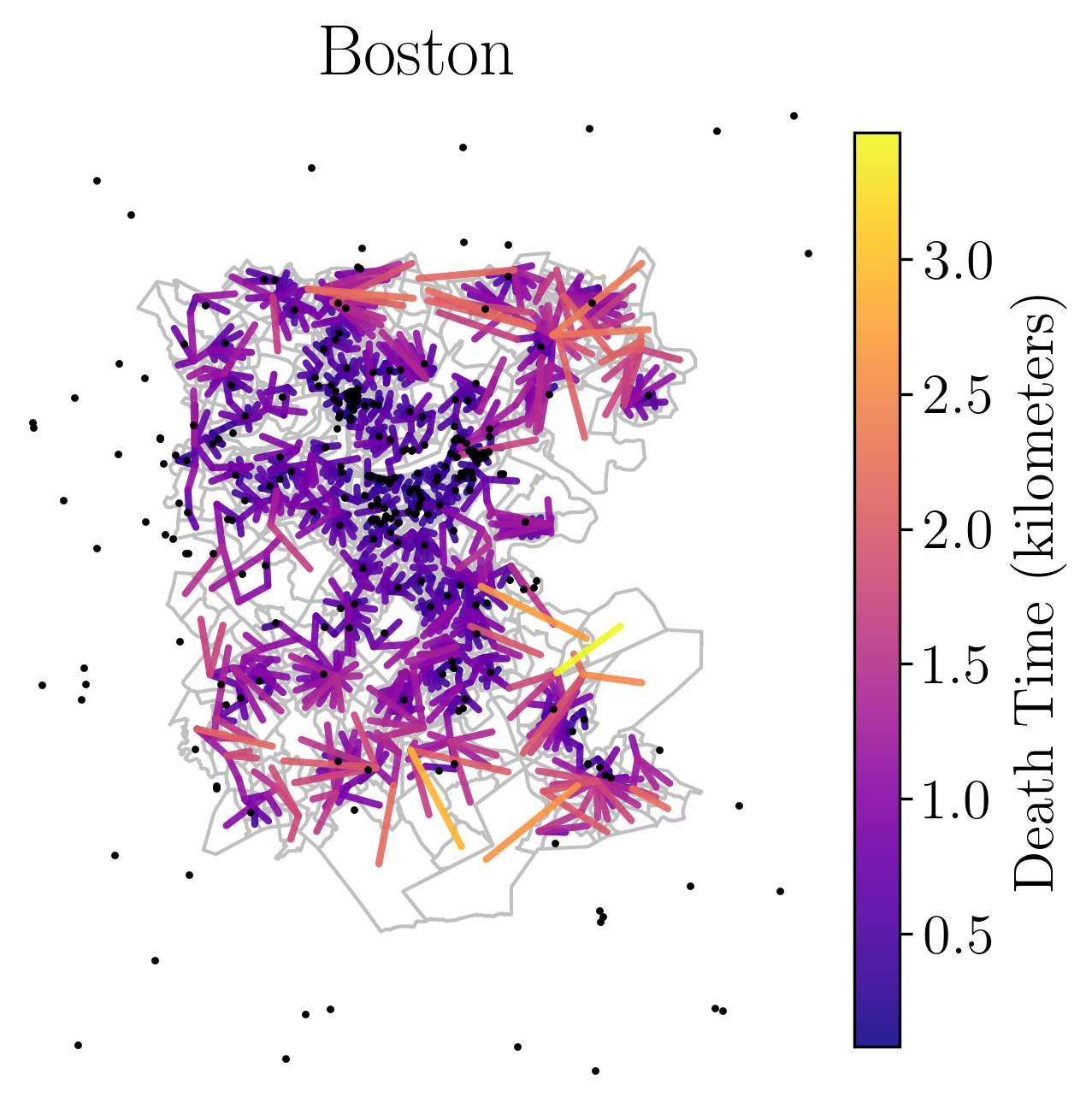} \\
\includegraphics[width=0.48\textwidth]{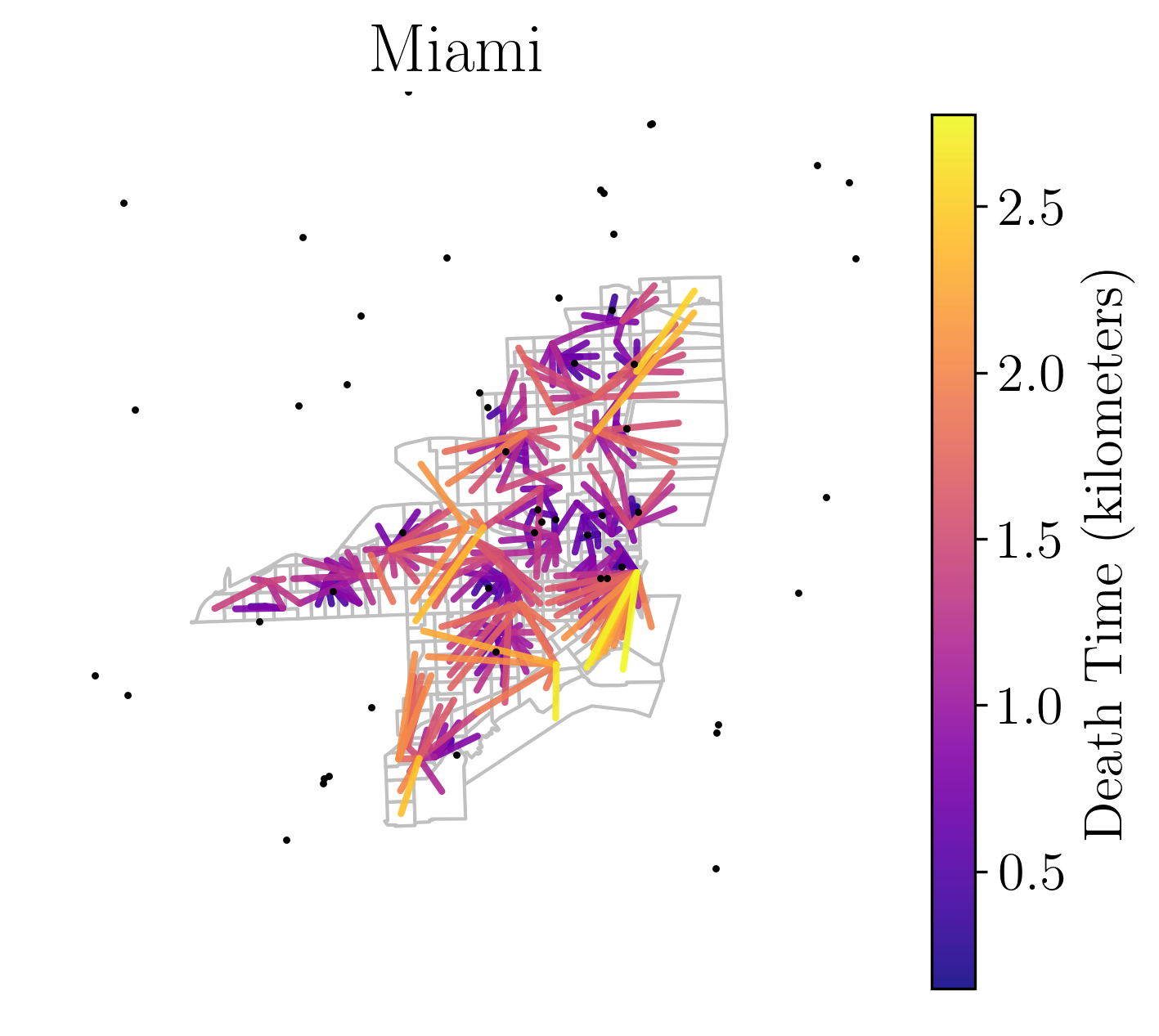}
\includegraphics[width=0.48\textwidth]{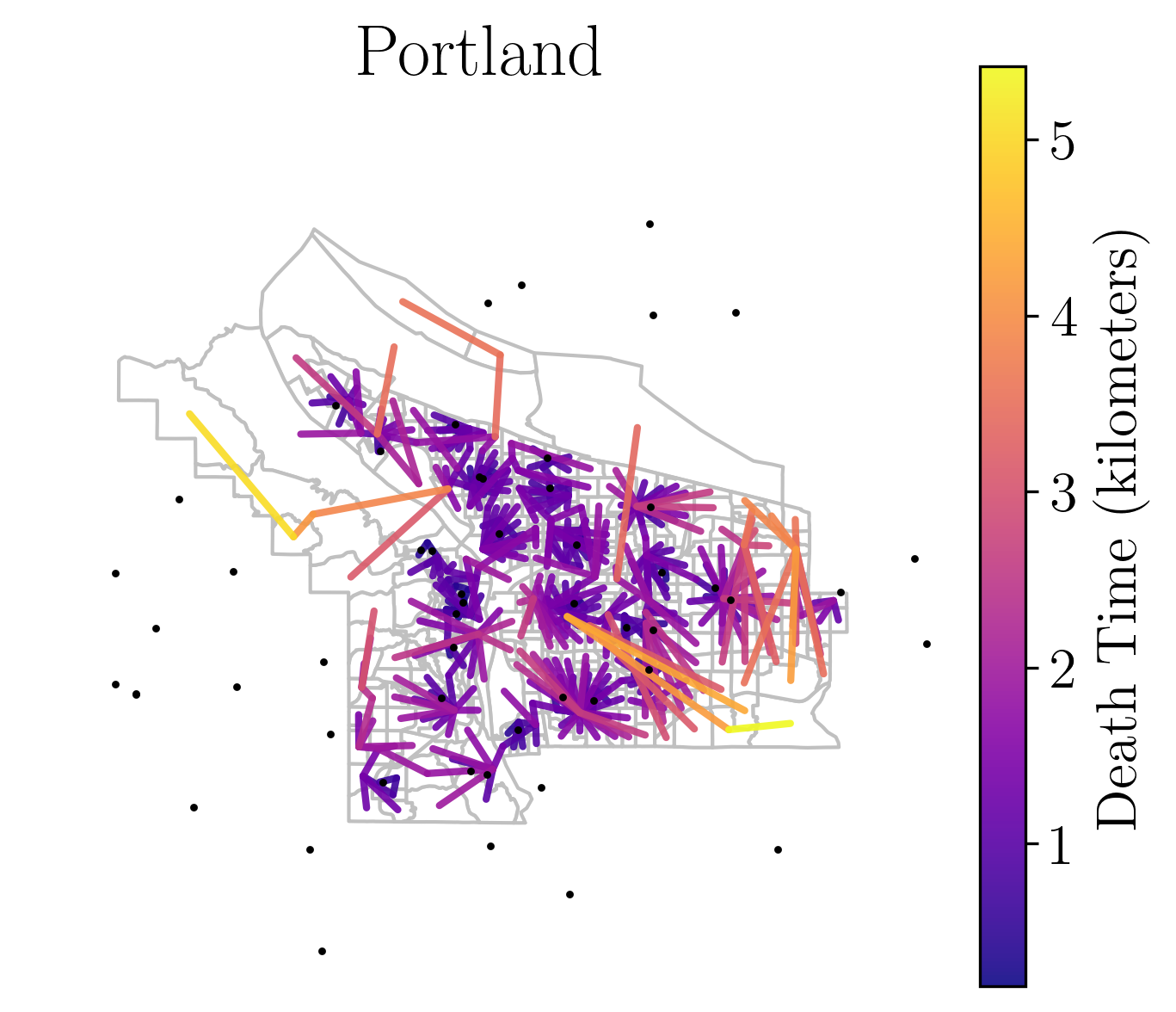}
\caption{Death simplices for the 0-dimensional homology classes.
The black dots represent the locations of cooling centers, including those outside the city but in neighboring areas. 
Maps are not to scale and the scale of death values are different across the cities.} 
\label{0D}
\end{figure}

\subsubsection{Death Simplices}

\begin{figure}
\centering 
\includegraphics[width=0.48\textwidth]{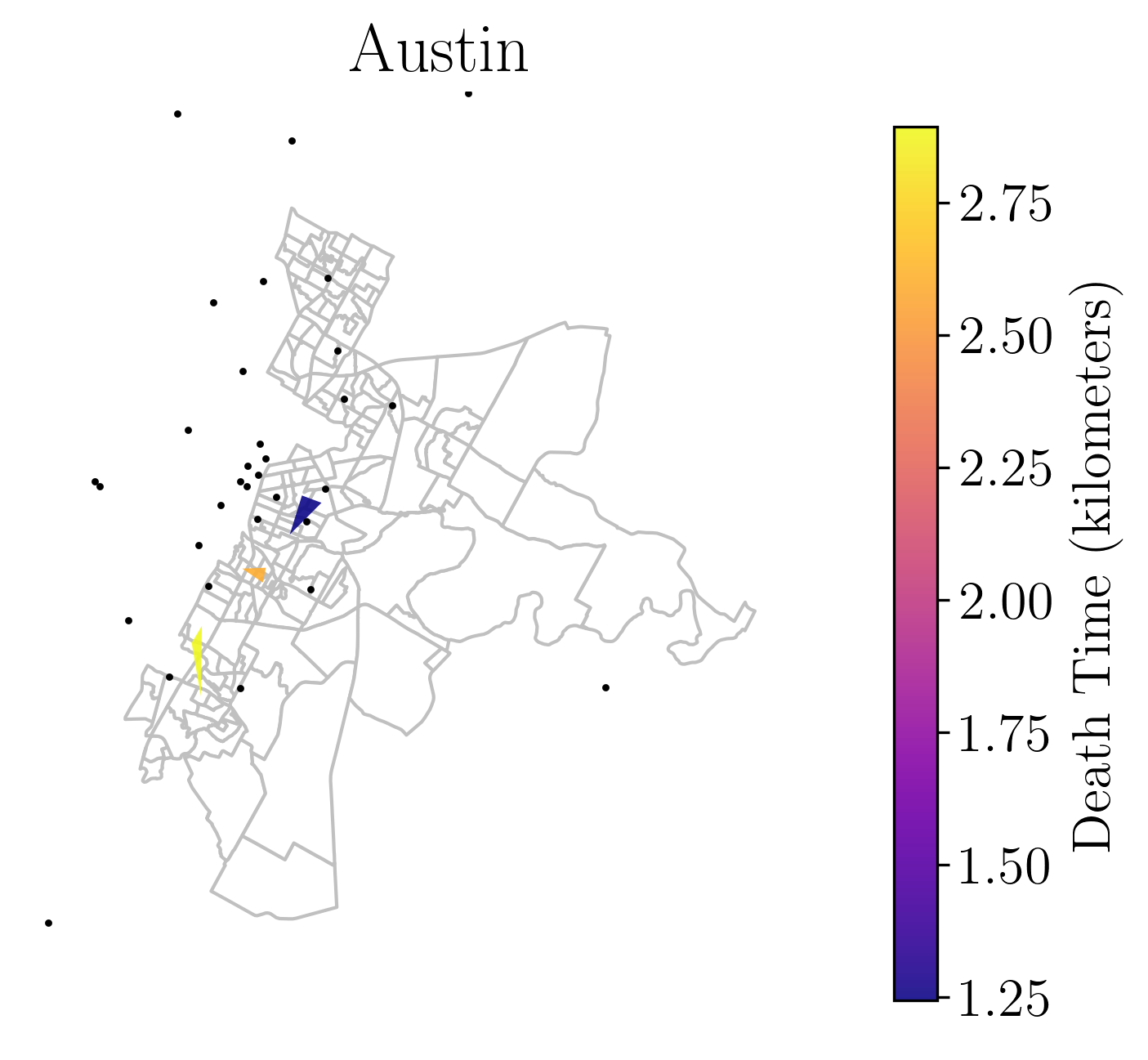}
\includegraphics[width=0.45\textwidth]{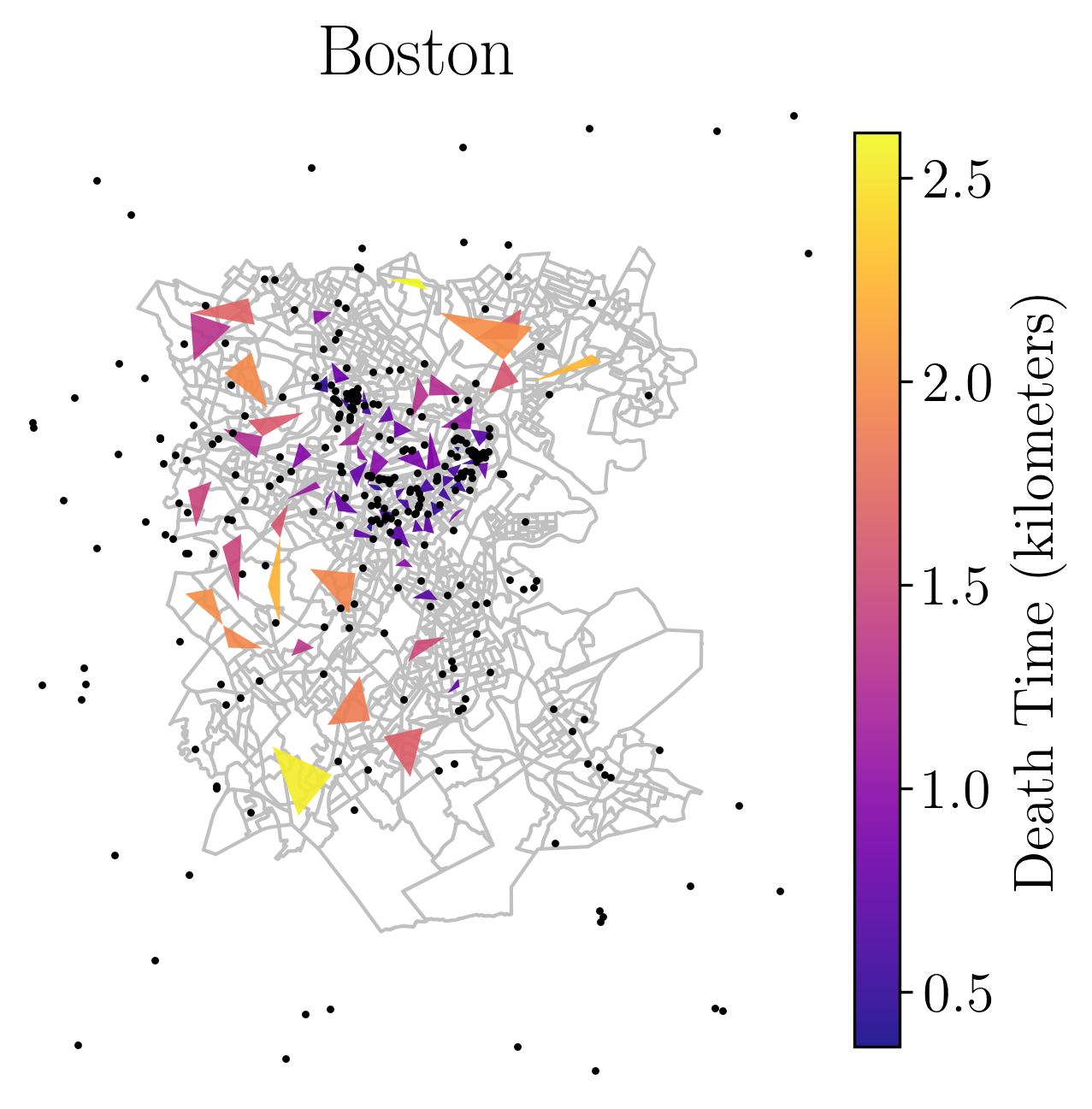} \\
\includegraphics[width=0.48\textwidth]{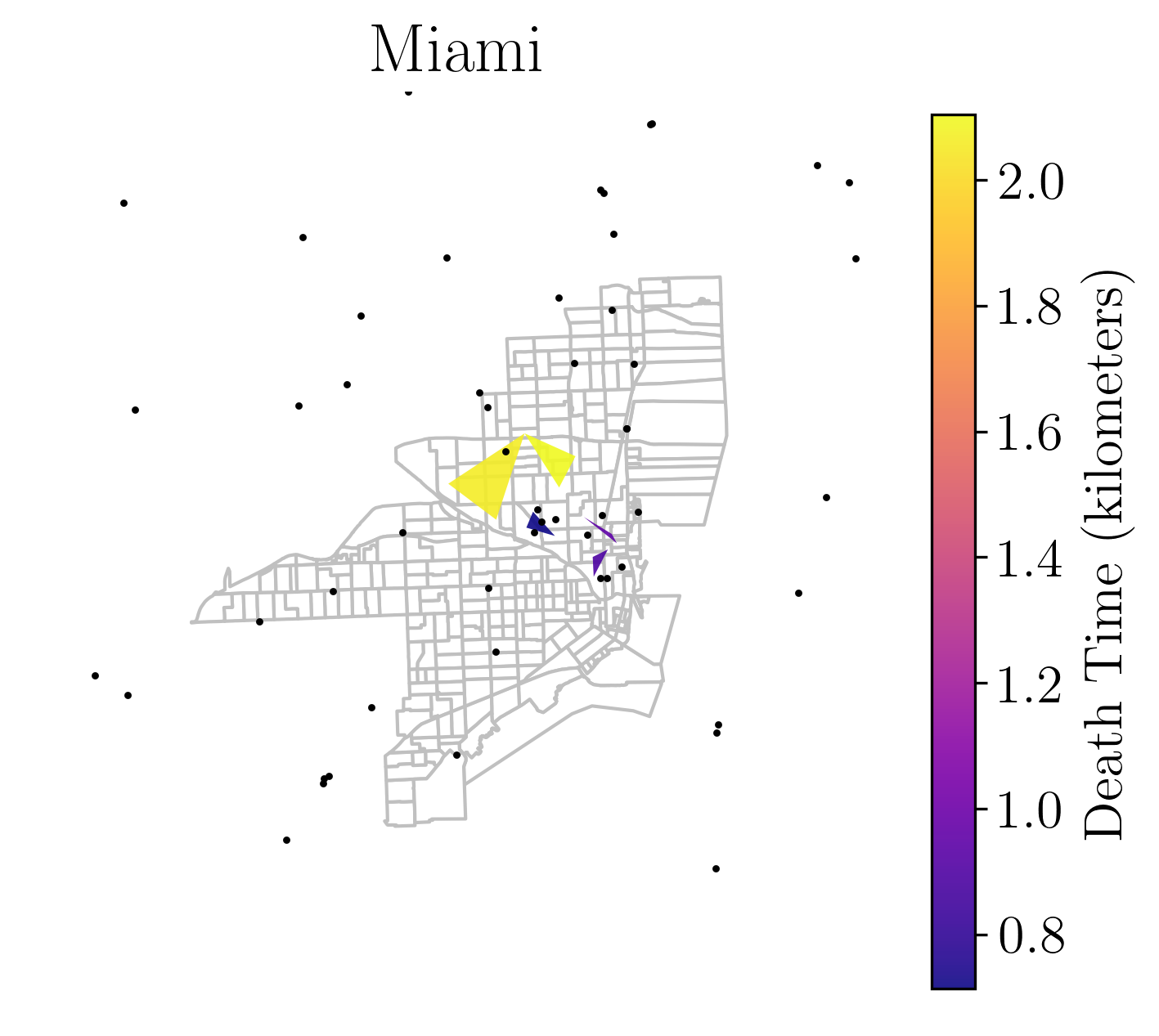}
\includegraphics[width=0.48\textwidth]{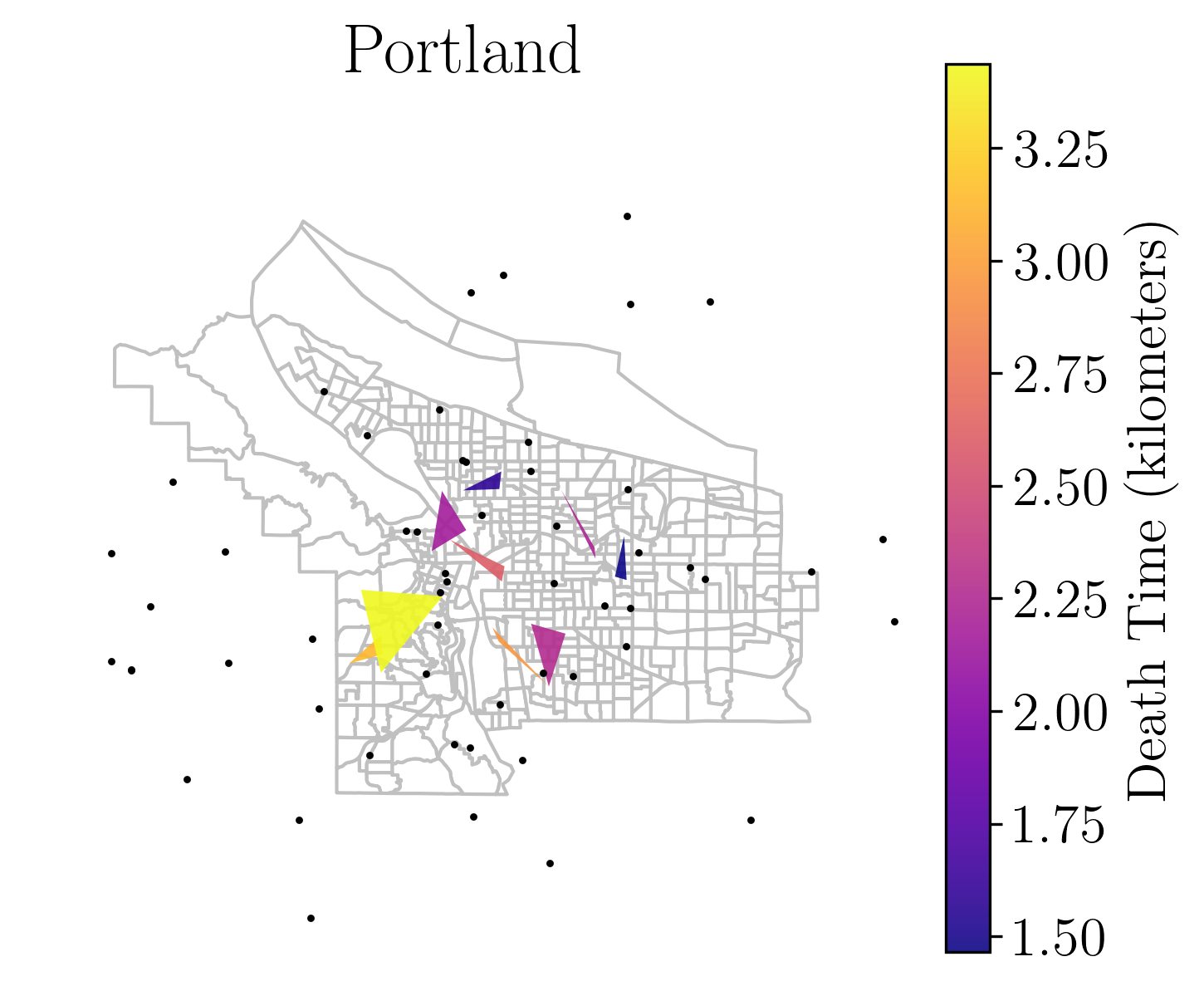} 
\caption{Death simplices for the 1-dimensional homology classes. 
The black dots represent the locations of cooling centers, including those outside the city but in neighboring areas. 
Largest death values are represented in yellow. Maps are not to scale.}
\label{1D}
\end{figure}

As discussed in Section~\ref{ssec:deathsimps}, we use the death simplices to visualize the regions of vulnerability.
Figures~\ref{0D} and \ref{1D} map the death simplices of the 0- and 1-dimensional homology features respectively.

In dimension 0, death simplices indicate a stretch of space along which there is a lack of cooling center coverage. 
Those with larger death values are interpreted as connecting one region to one with higher risk of heat-related mortality. For instance, in Austin, the edge connecting the southernmost census block to the center of the city has an especially large death time. 
Similarly, there are several death simplices with a high death time on the eastern side of the city.
This indicates that the southern and eastern portions of the cities lack coverage and are therefore those who reside in those areas are vulnerable to extreme heat. 
Other death simplices with high death times can be observed in the south-east area of central Boston, on the northern tip and south-eastern edge of Miami, and the north-west and south-east tip of Portland.

In dimension 1, the death simplices indicate the epicenter of the hole in coverage of cooling centers.
In Figure~\ref{1D}, we see one or more death simplices with a high death time in the southern regions of Boston and Portland as well as in central Miami. 
These regions are the most vulnerable to heat related mortality.
Boston also has numerous death simplices with greater-than-average death values.
This indicates Boston has more notable holes in coverage than the other three cities.

\begin{figure}[!ht]
    \centering
    \includegraphics[width=\linewidth]{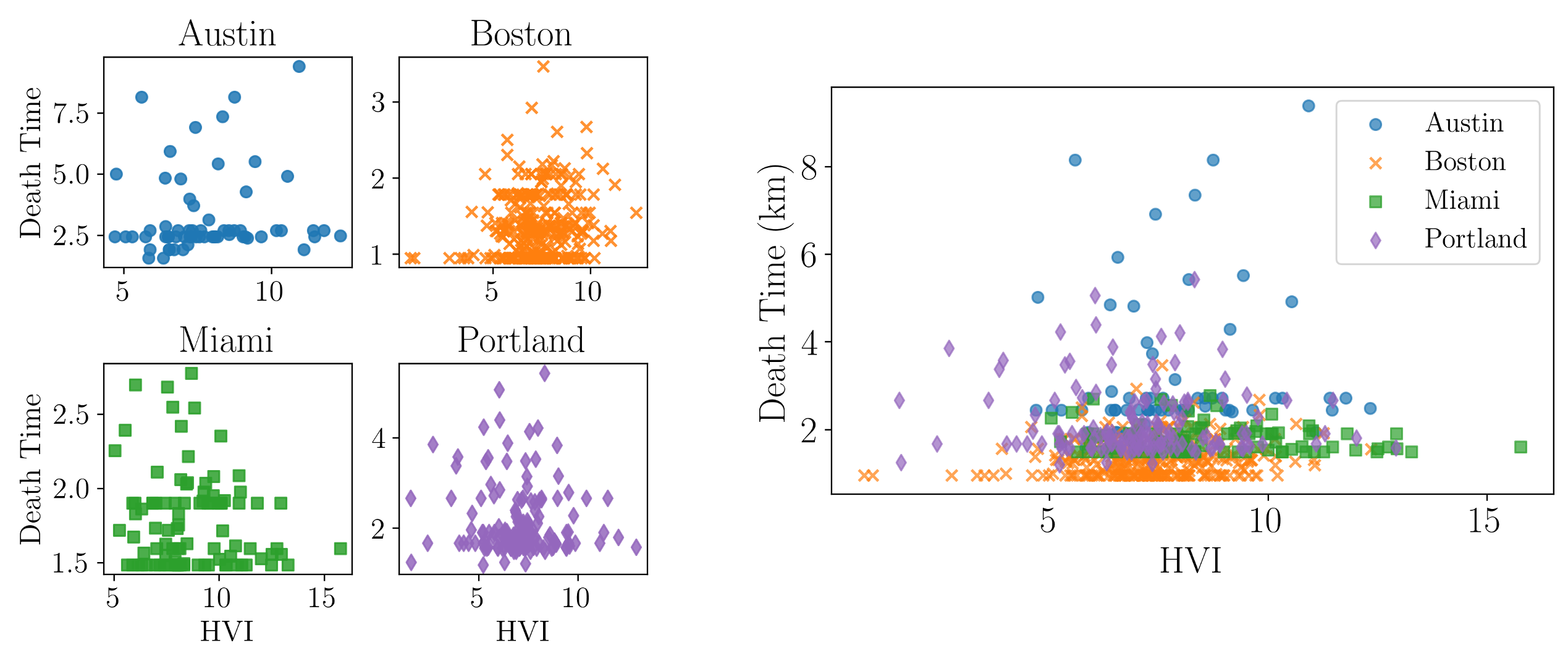}
    \caption{Comparison of the two measures of vulnerability. The vertical axis plots the maximum death time of components containing any census block within a given census tract. The horizontal axis plots the corresponding HVI for the census tract.
    Table~\ref{tab:correlations} shows the correlation and corresponding $p$-value for each city individually as well as all four combined.}
    \label{fig:scattercompare}
\end{figure}

\begin{figure}
\centering 
\includegraphics[width=0.85\textwidth]{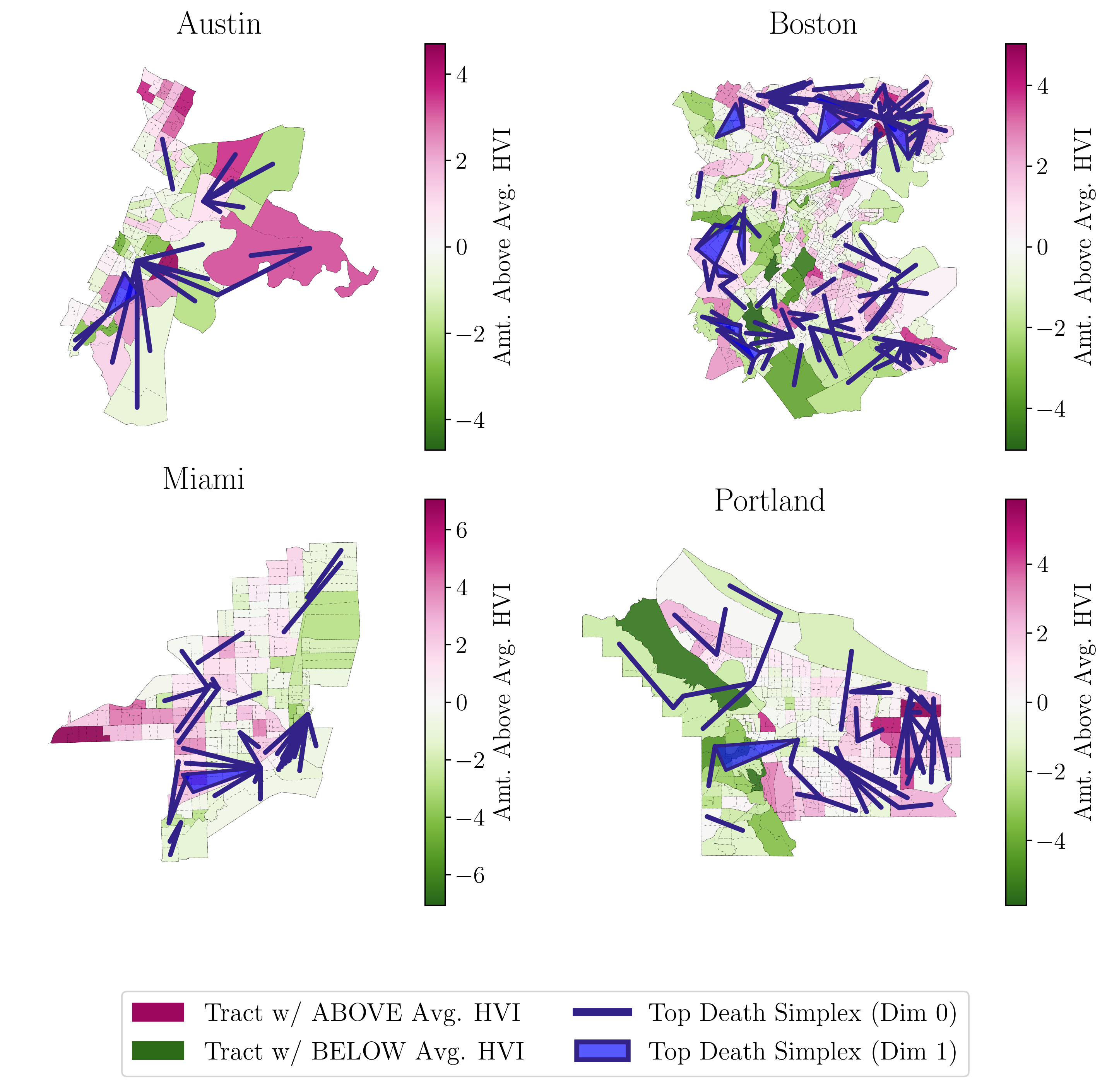}
\caption{In each map, census tracts are colored based on their HVI value relative to the average HVI within the city. 
Therefore pink and green tracts have an HVI above and below average, respectively. 
The blue lines and triangles are the death simplices for dimension 0 and 1 with death times in the 90\% percentile of death times within that city. 
}
\label{fig:comparevulnerable}
\end{figure}

\subsection{Comparing Approaches} \label{ssec:compareresults}

The HVI and topological approaches each identify regions of high vulnerability.
Here we compare the regions identified by each approach.

For a simple initial comparison, for each census tract we compute the maximum death time of any census block in that tract and plot it against the HVI of that tract in Figure~\ref{fig:scattercompare}.
It is hard to see any patterns visually, so we compute the correlations of each city individually as well as all four together.
These correlations are reported in Table~\ref{tab:correlations}.
Only Boston has a statistically significant correlation with a correlation coefficient of 0.132.
However, based on the large $p$-values across the other cities, we cannot draw any significant conclusions.

Next we want to look at specific regions.
To do so, we plot the death simplices with a death time in the 90-th percentile and the HVI data on the same plot in Figure~\ref{fig:comparevulnerable}.
It is important to note that the colorbar in Figure~\ref{fig:comparevulnerable} does not represent the HVI.
Instead, for a given census tract $T$, it plots $HVI(T)-\text{mean}\{HVI(t) : t\in \mathcal{T}\}$ where $\mathcal{T}$ is the set of all census tracts within a given city.
Therefore the colorbars should be interpreted as a measure of how vulnerable a tract is relative to the average.
Census tracts colored pink (which corresponds to positive values) indicate census tracts with an HVI above the mean.
Therefore, any intersection between a death simplex and a pink (above average HVI) census tract indicate regions of agreement between the two measures of vulnerability. 
We will focus on analyzing this plot to compare when the two methods align and analyze when they differ.

In Austin, we see overlap of the death simplices for dimension 0 with the regions of above-average HVI.
In particular, on the northeastern portion of Austin, there are two census tracts with well above average HVI that also intersect with death simplices.
This indicates agreement between the approaches meaning the regions with high HVI also lack cooling center coverage. 
Because both methodologies agree in these locations, we can interpret these locations as being more severely vulnerable than the locations identified by only one approach.

In Boston, the northeastern and southeastern corners of the city contains numerous death simplices and a collection of census tracts with above average HVI, indicating regions of high vulnerability. 
Similarly, in eastern Portland there are several regions that are identified as vulnerable by both approaches, indicating another area of vulnerability.

We can also use this plot to identify regions of disagreement where one approach may identify the region as vulnerable while the other does not. 
These regions would indicate a moderate level of vulnerability.
For example, regions with above average HVI but no death simplices could be due to a high vulnerable population but in an area that is well covered by cooling centers.
Regions with a below average HVI but several death simplices could indicate areas that have lower temperatures and more tree cover but do not have many nearby cooling centers.
An HVI well below average with death simplices might indicate a low-population area where cooling centers are therefore not as necessary.
We see examples of all of these in Figure~\ref{fig:comparevulnerable}.

For example, in Figure~\ref{0D}, we see one death simplex in Boston with a particularly high death time on the eastern side of the city.
Based on this we would interpret this region as being highly vulnerable, however this particular census block is located on an island in Boston Harbor called Moon Island.
It was previously a sewage treatment plant but now is used as a training facility for the Boston fire and police departments \cite{moonisland}.
While it is true this area has poor access to cooling centers, it has a below average HVI, indicating it is not an area of concern.  
We can investigate further and see in Figure~\ref{fig:HVI_figs_all} that the census tract has a very low population of individuals of age $<5$ or $>65$.
Similarly, in northwest Portland, there is a death simplex with high death time but the corresponding area has a low HVI due to a low number of vulnerable individuals.

Conversely, in south Boston, there are a collection of census tracts with lower-than-average HVI but where there are several death simplices in the area. 
Consulting Figure~\ref{fig:HVI_figs_all}, we see the area has low average temperatures and a high percentage of tree cover, indicating lower vulnerability, but there is a fairly high number of individuals of age $<5$ or $>65$.
So while the HVI would not indicate this as an area of high priority, the lack of cooling centers makes it more concerning.

\section{Conclusion and Discussion} \label{sec:conclusions}

In this paper, we use persistent homology to supplement existing approaches for identifying areas with high risk of heat related mortality.
In particular, we develop an approach for identifying locations in need of cooling centers to reduce mortality risk.
When analyzed together with an HVI, a standard approach, we see a more comprehensive understanding of heat-related vulnerability.

Our TDA approach only requires the current distribution of cooling centers and a collection of landmarks (e.g. the centroids of census tracts or census blocks) within a given city. 
Demographic data, which can be difficult to collect, is not needed. 
However, when used independent of the HVI, we see the topological approach may over or underestimate the vulnerability of a given area.
Thus TDA is not a replacement to standard techniques such as an HVI map, but it can identify additional high-risk locations and aid in prioritizing existing high-risk areas. 
Our topological approach captures the spatial distribution of cooling centers and considers neighboring cities, adjacent blocks, and the proximity of those blocks to cooling centers. 
In conjunction with the HVI, it provides a more holistic evaluation of the vulnerability than can be identified using either approach individually. 

Further, our methodology is adaptable and can be used to study access to other resources across various geographic scales other than at the census block level. We address potential future topics of study at various geographic scales in Section \ref{future}.

\subsection{Limitations} 

Our methodology relies on a dataset of cooling center locations using OpenStreetMap (OSM), details of its creation are presented in Appendix~\ref{appendix:data cooling centers}. 
However, it is important to recognize that our analysis is dependent on the quality of OSM data. 
For instance, the map itself may have incomplete data about a city's current distributions of cooling centers, which could impact our results. 
In addition, our methodology does not account for a cooling center's hours of operation nor capacity. 

Similarly, we are limited by the completeness of the demographic data. 
For instance, we restricted our attention to central Austin, TX and central Boston, MA because the U.S. Urban Heat Island Mapping Campaign did not have comprehensive data for the full city. 
In addition, ideally we would have demographic data at the census block level rather than the census tract level.

Further, as discussed in Section~\ref{ssec:distance_computation}, we use the geodesic distance between census blocks and cooling centers.
While some work suggests this is an appropriate proxy for travel distance, it cannot account for variables that influence travel such as traffic and road closures.
Ideally we would like to use travel time like the approach in \cite{Voting}, however google maps has a fee for each distance calculated, so using travel time or different modes of transportation was infeasible in this case.

It is also important to address the topographical features that may limit our analysis. 
For one, census blocks can be different sizes within a given city. 
This is a problem, for example, because a centroid in a large census block may appear to be isolated spatially in our analysis and, therefore, not connect to other centroids until later in the filtration. 
This census block could be labeled as not being well covered, when in fact it is. 
Ideally, a census block's centroid would be a good geographical approximation for any neighborhood within the block. 
A future improvement could include sampling multiple points within a census block rather than only using the centroid.
In addition, our analysis does not account for bodies of water. 
For example, some unpopulated islands off the coast of Boston are included within census block boundaries without regards to the ocean.

Lastly, we note again that death simplices are not stable. Thus, small changes in the data could change the location of the death simplex, leading to different conclusions.

\begin{table}[ht]
\caption{Correlation between death time and HVI, as shown in Figure~\ref{fig:scattercompare}. The $p$-value is computed from a two-sided $t$-test. The last column labeled ``All'' indicates the correlation if we combined the points for all four cities. Note that Boston is the only city with a $p$-value $<0.05$ with a correlation of 0.132, indicating a slightly positive correlation.}
    \centering
    \begin{tabular}{l c c c c c}
        \hline\hline
         & \textbf{Austin} & \textbf{Boston} & \textbf{Miami} & \textbf{Portland} & \textbf{All} \\
        \hline
        Correlation Coefficient & 0.096 & 0.132 & -0.140 & -0.075 & 0.048 \\
        p-value & 0.446 & 0.018 & 0.171 & 0.372 & 0.234 \\
        \hline
\end{tabular}
    \label{tab:correlations}
\end{table}

\subsection{Future Work}\label{future}

In the future, we would like to improve our method by changing how we capture the vulnerable regions from the topological approach. 
As discussed, death simplices are not stable which can lead to different results with small changes in the input data.
Using minimal cycle representatives \cite{li_minimalcyclerepresentatives_2021} to identify the holes in coverage could avoid this issue and allow for better identification of areas with poor access to resources.
However, this raises numerous other decisions and concerns as minimal cycle representatives are not uniquely defined and can be challenging to compute. 

We would also like to explore other spatially aware approaches for detecting heat vulnerability from the literature as additional points of comparison.
We'd also like to conduct a larger scale investigation about additional cities.
We chose four different cities to test our method, however there are many other cities included in the US Urban Heat Island Mapping Campaign. 

Our methodology could be used in many other applications as well; for example, studying the coverage of resources such as food banks or healthy food options.
On a smaller scale, one could consider the locations of automated external defibrillators (AEDs) or fire extinguishers within a large building, such as a school.
A resource that will be important over the coming decades is electric vehicle charging stations. 
For this application, a larger statewide or nationwide scale would be useful to determine areas where the addition of a charging station would facilitate the use of electric vehicles in the area.

\section*{Acknowledgments}
We thank Mason Porter and Nicole Sanderson for helpful comments and discussions. 
We would also like to thank the two anonymous reviewers for their constructive comments and suggestions, which helped improve the paper.
EO would like to acknowledge the support and funding provided by UCLA's Queen's Road Foundation. 
Lastly, we thank the developers of the \texttt{GeoPandas} \cite{geopandas}, \texttt{Shapely} \cite{shapely2007}, \texttt{networkx} \cite{networkx}, and \texttt{GUDHI} packages \cite{gudhi} for their open source code used for the creation of our maps, analysis, and figures.

\appendix

\section{Data Collection} \label{AppendixData}
\begin{figure}
\centering 
\includegraphics[width = 0.7\textwidth]{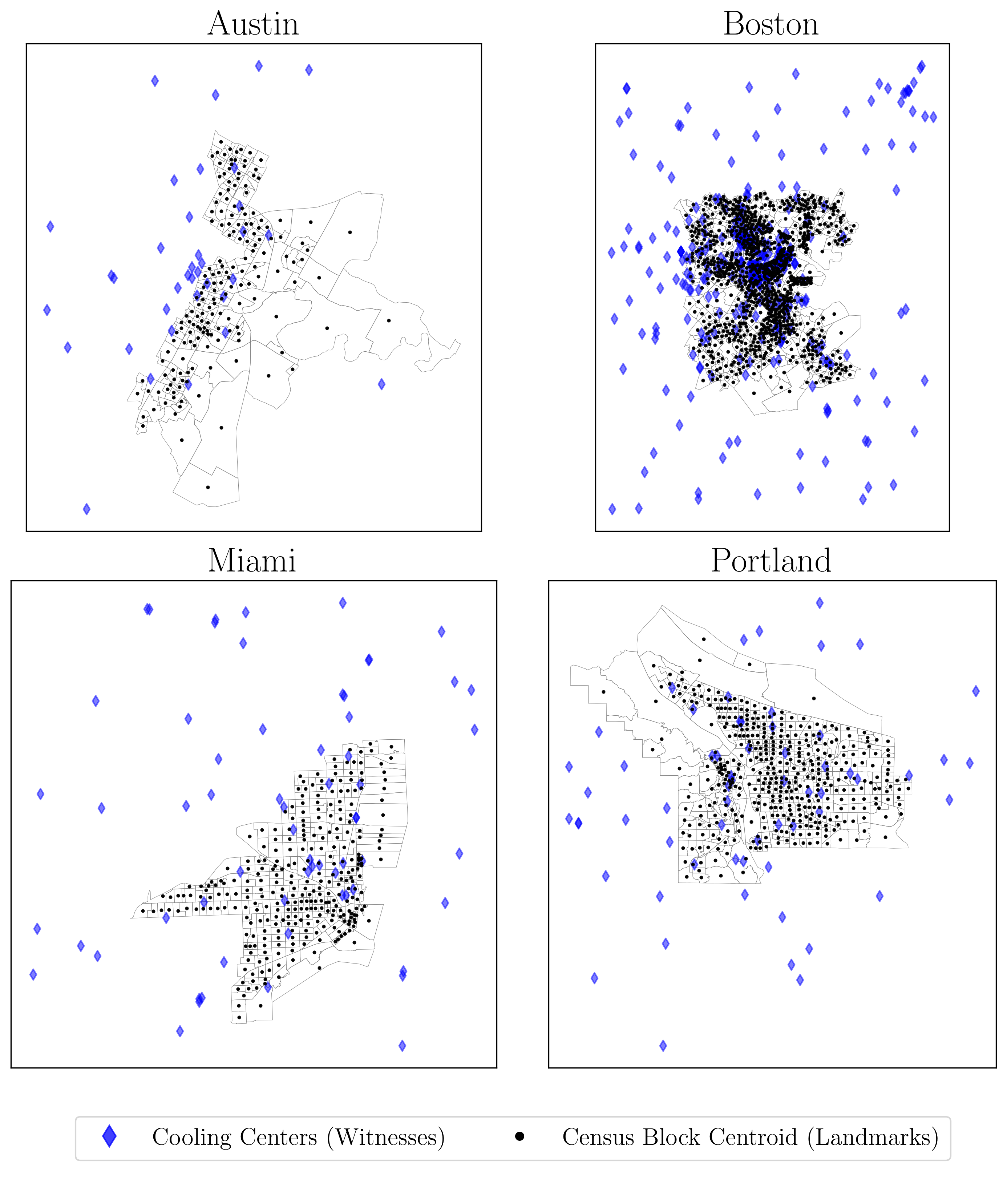}
\caption{Final dataset of census tract centroids (landmarks) and cooling centers (witnesses) for each of the four locations, indicated as black dots and blue diamonds, respectively.}
\label{data}
\end{figure}

In this section we describe the data sources along with any pre-processing steps applied to them. 
Section~\ref{appendix:data centroids} describes the landmark (census block centroids) data, Section~\ref{appendix:data cooling centers} describes the witness (cooling center) data, and Section~\ref{appendix:data score} describes the HVI data. 
Figure \ref{data} plots the full dataset of landmarks and witnesses.

\subsection{Landmark Data Collection and Processing} 
\label{appendix:data centroids}
Our approach requires a Shapefile containing 
some set of standardized geographic units that subdivide the region of interest into discrete, mappable areas.
There are many different scales of these geographic units. Three examples of regions based on data collected by the US Census Bureau \cite{census_defs} include: 
\begin{itemize}[itemsep=2pt,leftmargin=0.7cm]
    \item Census blocks: the smallest geographic division for which the US Census Bureau collects data. The population of each block is approximately equivalent to that of a city block in an urban area. 
    \item Census block groups: groups of contiguous census blocks containing a total of 600–3,000 people. 
    \item Census tracts: groups of contiguous census block groups containing a total of 2,500-8,000 residents.
\end{itemize}
For our analysis using persistent homology, we use census blocks as they allow for the most granular level of detail. 
However, one could use larger geographic units such as block groups or tracts.

The Shapefiles to map the census blocks were retrieved from Data.gov \cite{dataDatagovHome}. The following datasets were downloaded in July 2024: 
\begin{enumerate}[itemsep=2pt,leftmargin=0.7cm]
\item  TIGER/Line Shapefile, 2017, state, Massachusetts, Current Block Group State-based
\item TIGER/Line Shapefile, 2017, state, Florida, Current Block Group State-based
\item  TIGER/Line Shapefile, 2017, state, Oregon, Current Block Group State-based
\item TIGER/Line Shapefile, 2017, state, Texas, Current Block Group State-based
\end{enumerate}
From the Shapefile, we compute the centroid (the geographic center) of each census block using the \texttt{.centroid} attribute of the \texttt{Shapely} library in Python.

\subsection{Witness Data Collection and Processing} 
\label{appendix:data cooling centers}

We scan OpenStreetMap (OSM) \cite{OSM} to collect the latitude and longitude coordinates of cooling centers. In order for our methodology to accurately assess the coverage of cooling centers near the boundary of the city, we buffer our search area. A buffer is necessary because a census block on the edge of the city is not restricted to only visiting cooling centers within its city. In reality, residents of that census block could travel to nearby cooling centers of neighboring cities. To create the buffer, we find the longest distance between any two census tract centroids within the city and divide that distance by 2. We use this distance as a reference to buffer the city. 

We use a function within the OSMnx package, \texttt{geometries\_from\_bbox()}, to create the bounding box for our search area. The furthest southwest and northeast coordinate for the bounding box is summarized in Table \ref{table:coordinates}. We search for features with the following tags: ``library," ``community center," ``senior," and ``recreation center.'' These tags were selected based on the definition of a cooling center presented in \cite{Kim}. The dataset of cooling centers used in this study can be seen in Figure \ref{data} in dark green.

\begin{table}[htbp]
\label{table:coordinates} 
\caption{Furthest Southwest and Northeast coordinates used to search OpenStreetMap for cooling center tags.}
\begin{center}
\begin{tabular} { > {\centering\arraybackslash}m{2.5cm}   >{\centering\arraybackslash}m{4.5cm}  > {\centering\arraybackslash}m{4.5cm}  } 
  \hline\hline
  \textbf{City} & \textbf{Southwest Coordinate}  & \textbf{Northeast Coordinate}\\ 
  \hline\hline
  Austin, TX & (-97.90787741442931, 30.033924160573395) & (-97.55280269603111, 30.487698664291752)\\
  \hline
   Boston, MA & (-71.29861054113876, 42.10774101580344) & (-70.85728173592592, 42.547921903453464)\\
  \hline
  Miami, FL & (-80.38082496276087, 25.68912671238342) & (-80.09556728681095, 25.9273454531309)\\ 
  \hline
  Portland, OR & (-122.8805563576074, 45.28572461680384) & (-122.35668691377035, 45.69996159808512)\\ 
  \hline
\end{tabular}
\end{center}

\end{table}

\subsection{Heat Vulnerability Index Data Collection and Processing} 
\label{appendix:data score}
\begin{figure}
    \centering
    {\large Austin} \\
    \includegraphics[width=0.24\linewidth]{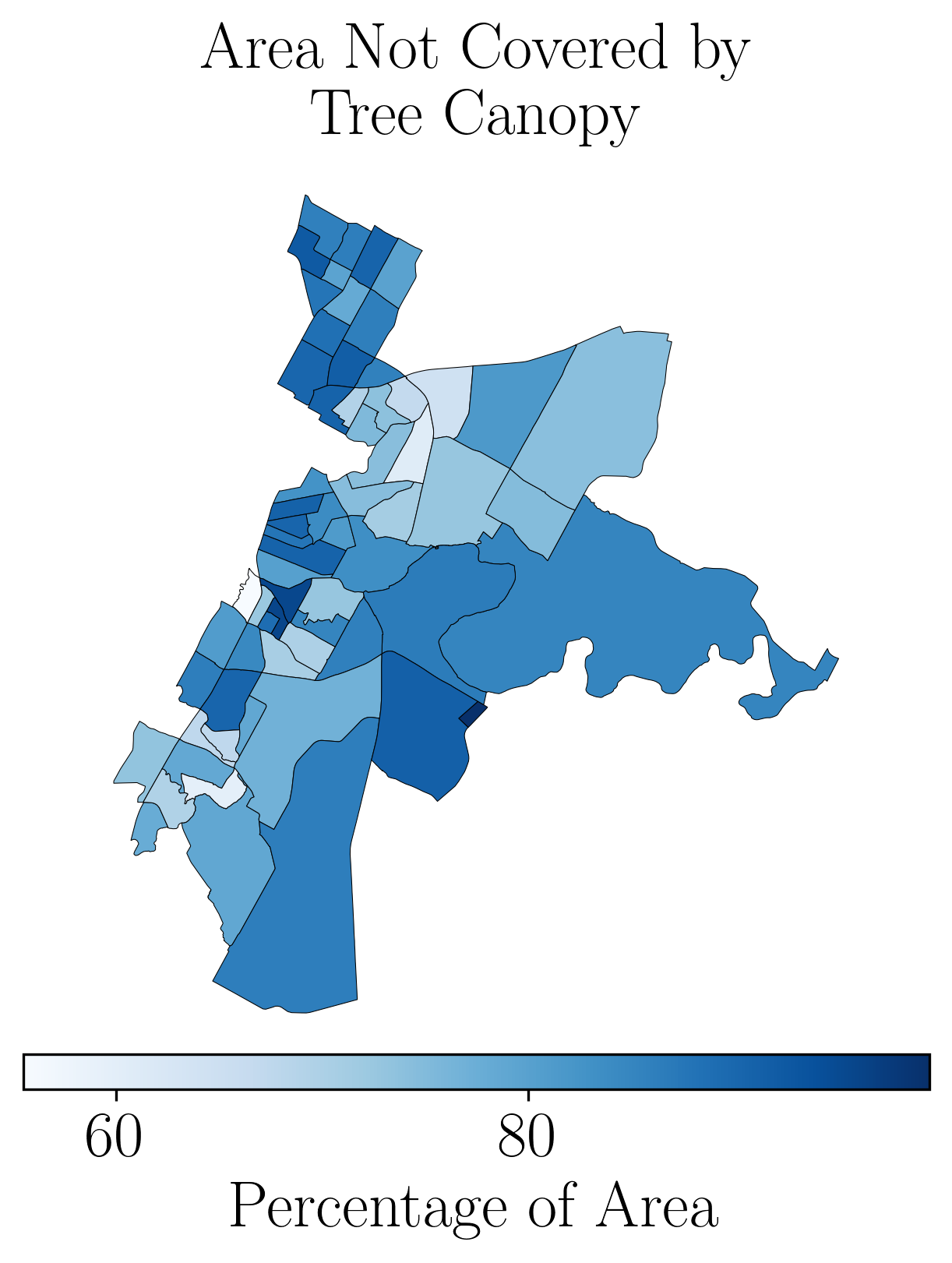} 
    \includegraphics[width=0.24\linewidth]{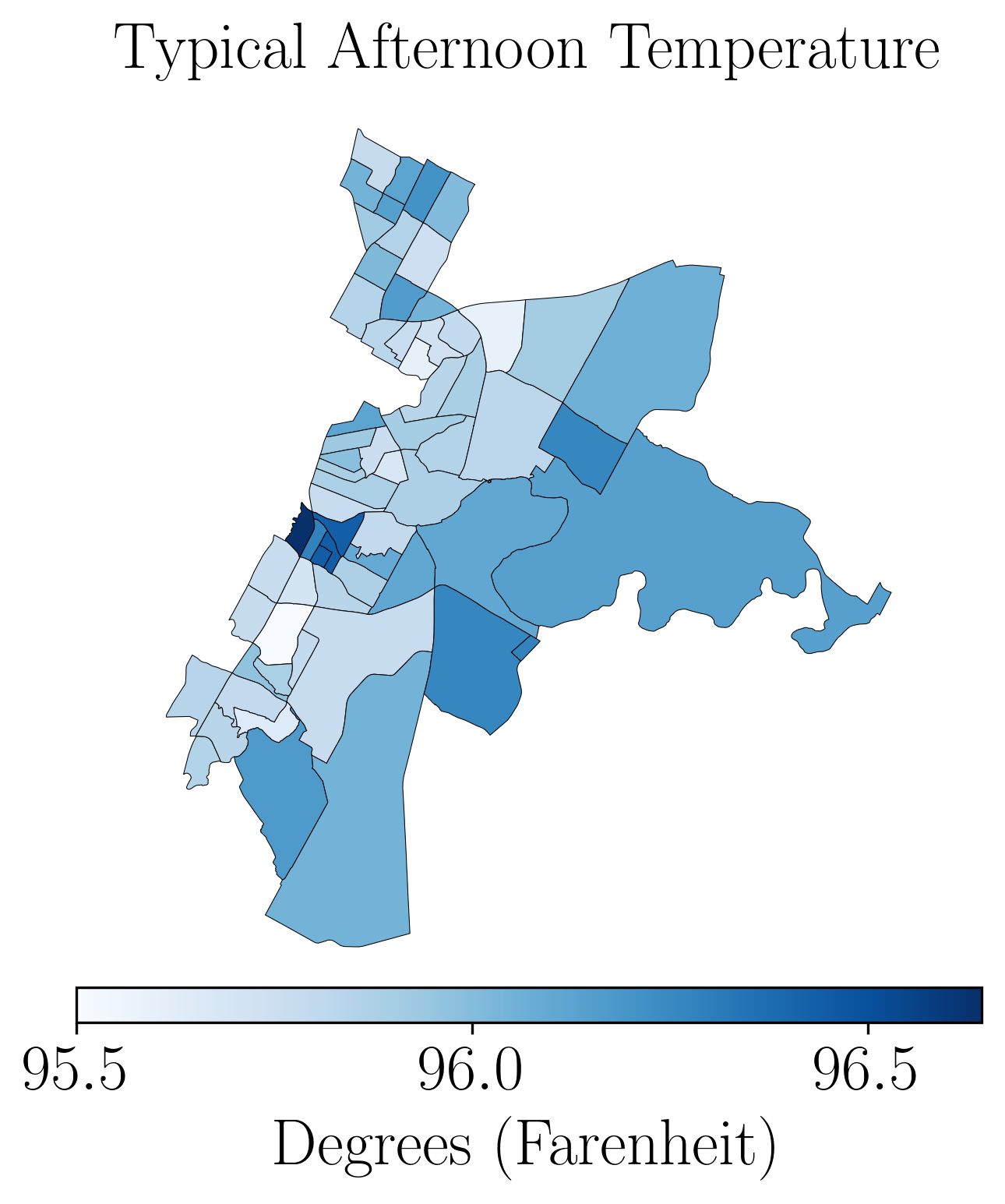}
    \includegraphics[width=0.24\linewidth]{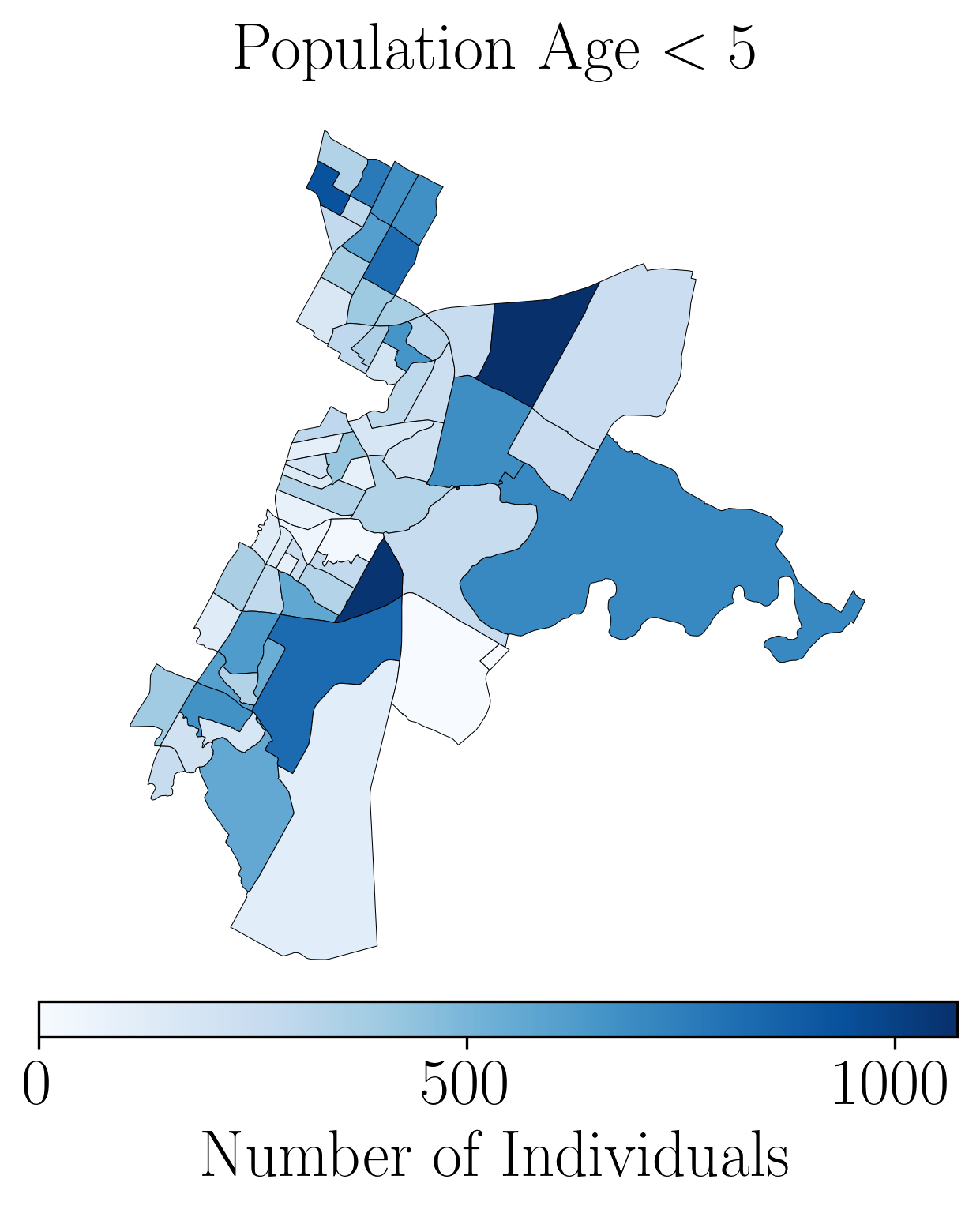} 
    \includegraphics[width=0.24\linewidth]{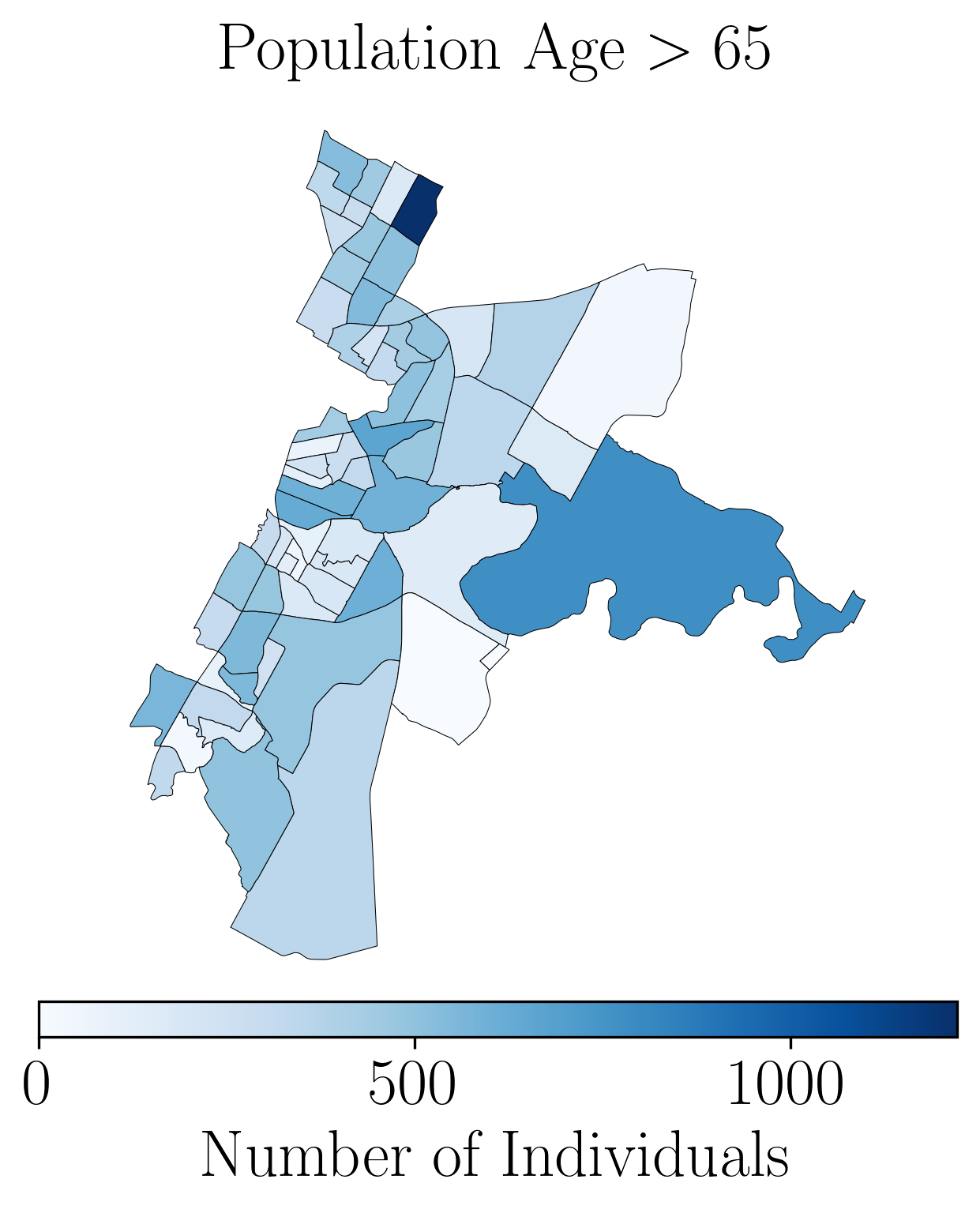} \\

\vspace{0.3cm}

    \centering
    {\large Boston} \\
    \includegraphics[width=0.24\linewidth]{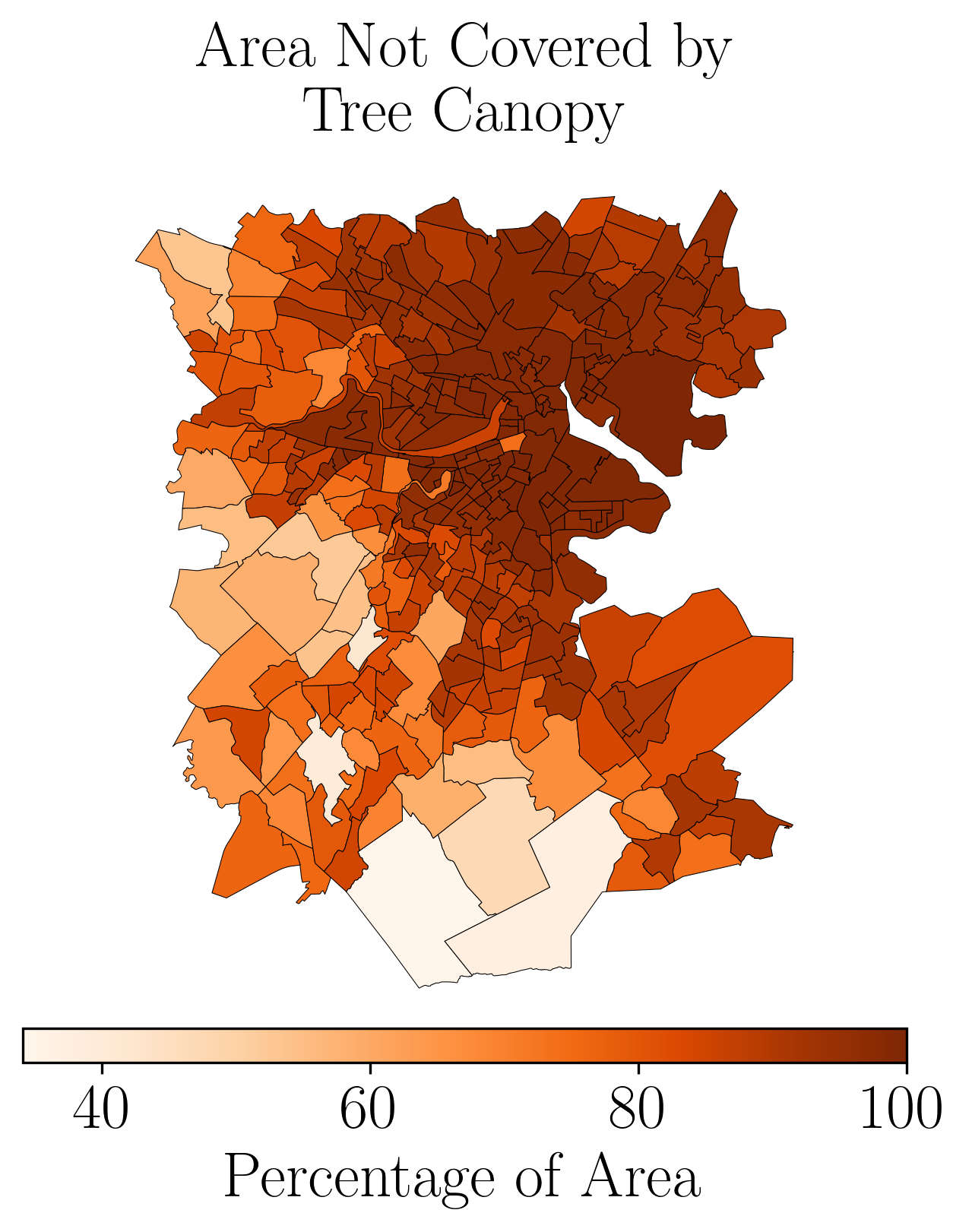}  
    \includegraphics[width=0.24\linewidth]{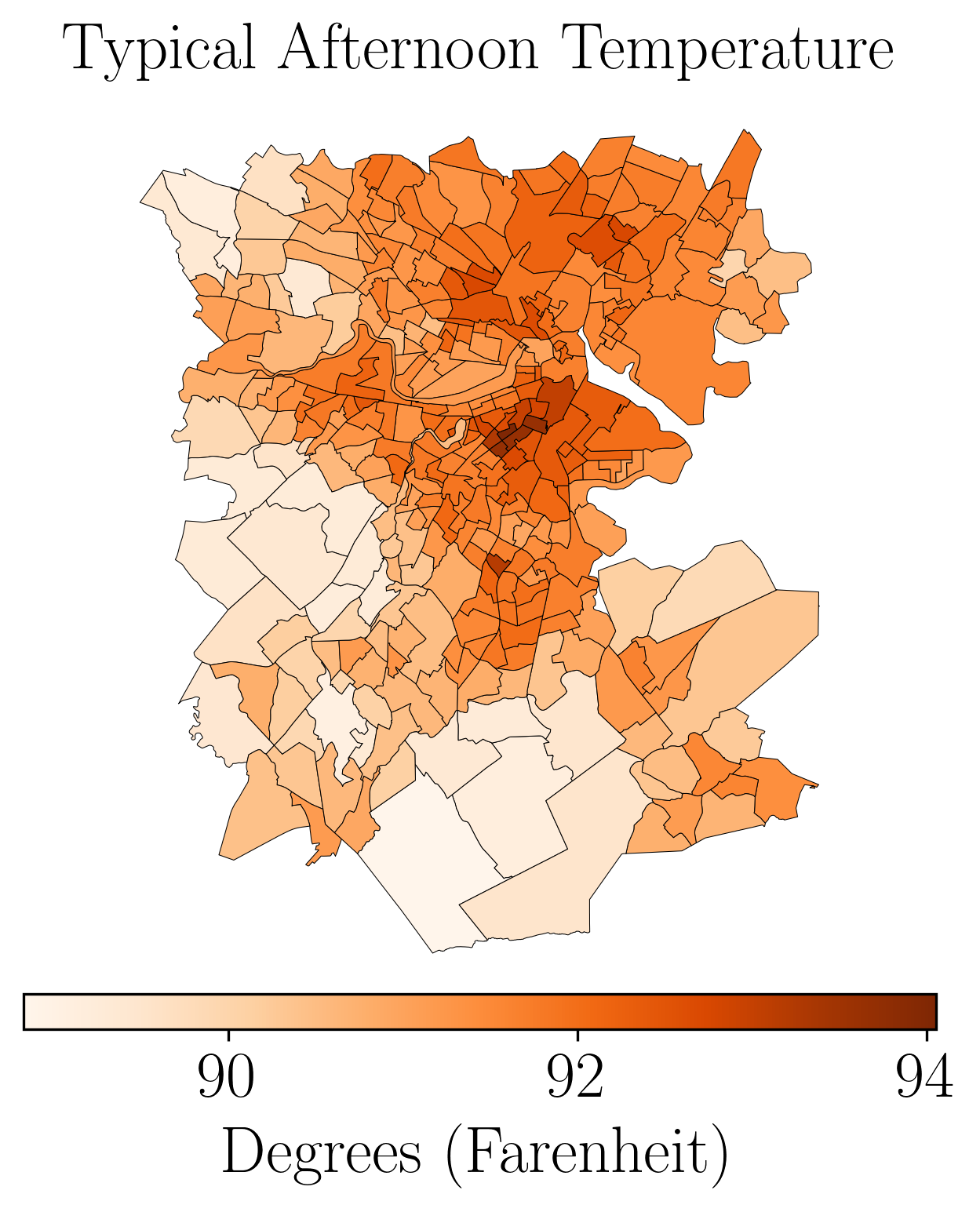}  % \\
    \includegraphics[width=0.24\linewidth]{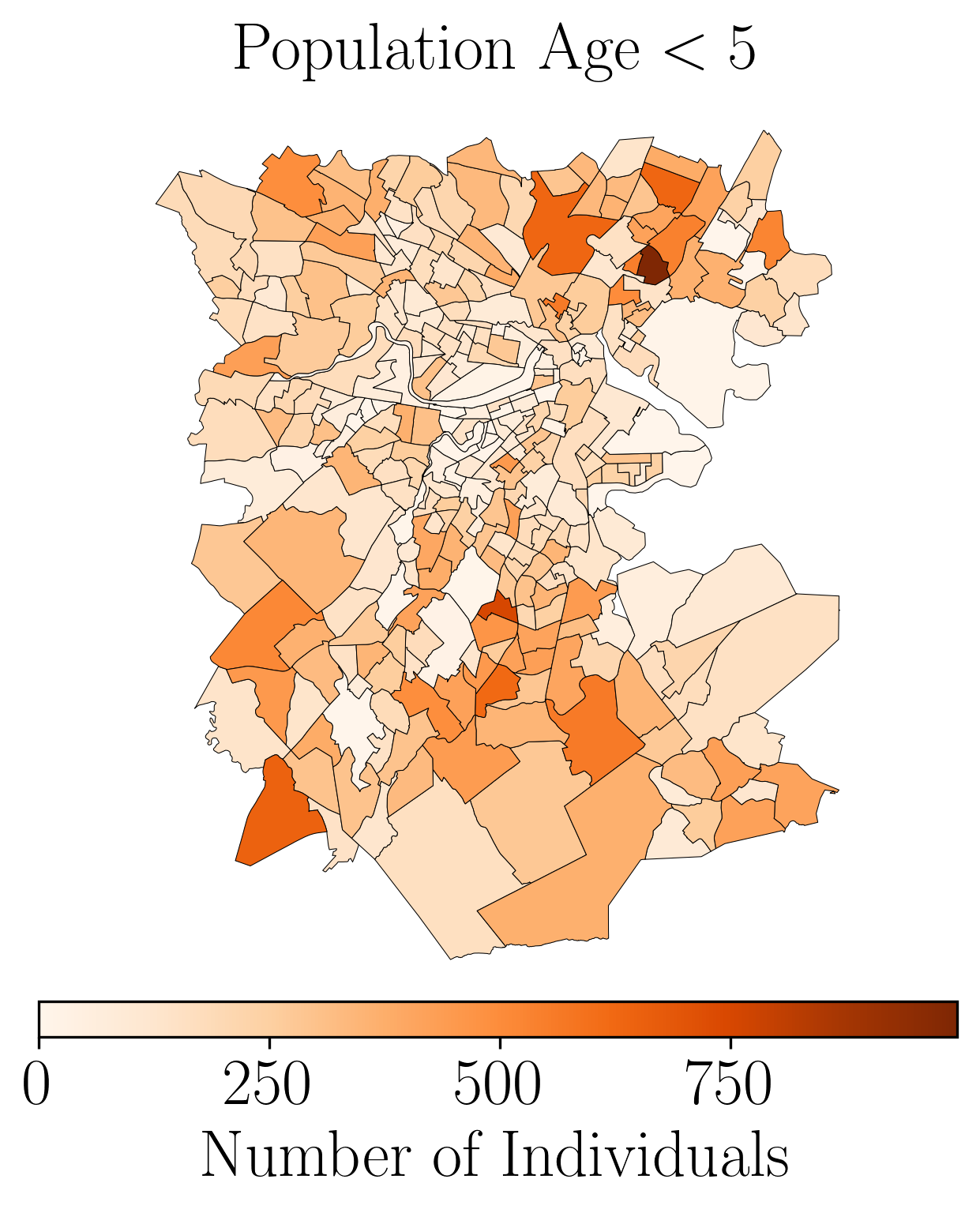} 
    \includegraphics[width=0.24\linewidth]{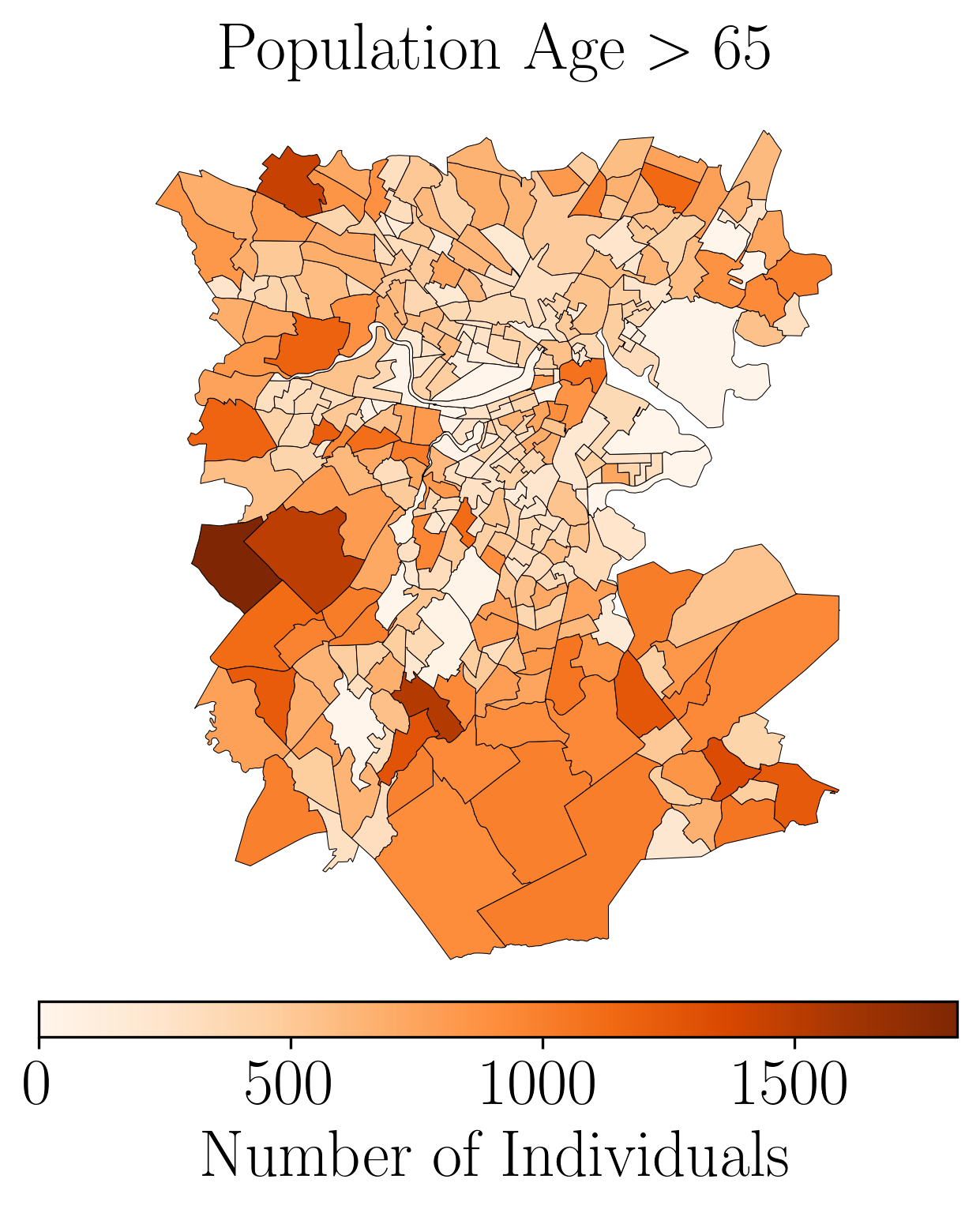}  \\

\vspace{0.3cm}

    \centering
    {\large Miami} \\
    \includegraphics[width=0.24\linewidth]{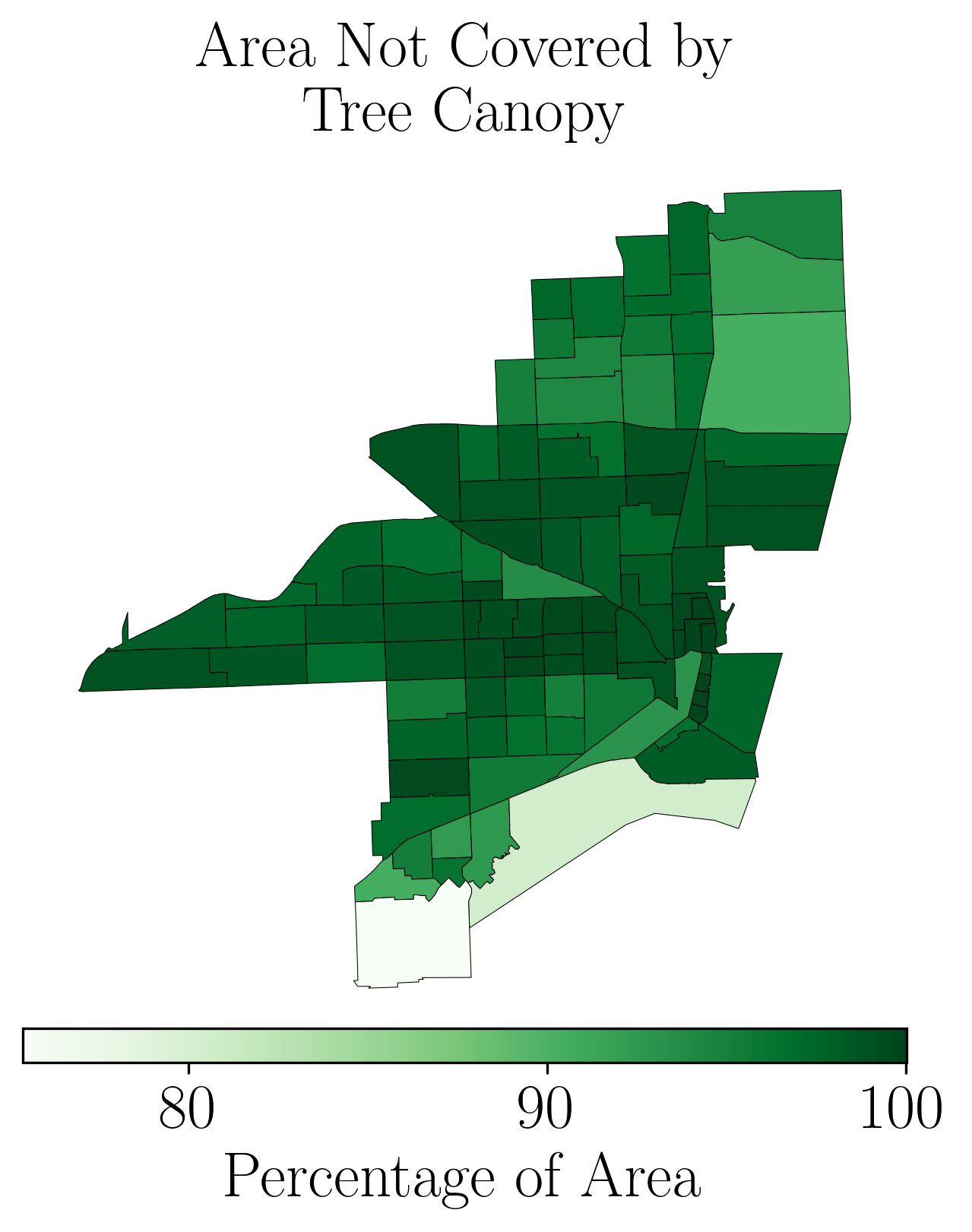} 
    \includegraphics[width=0.24\linewidth]{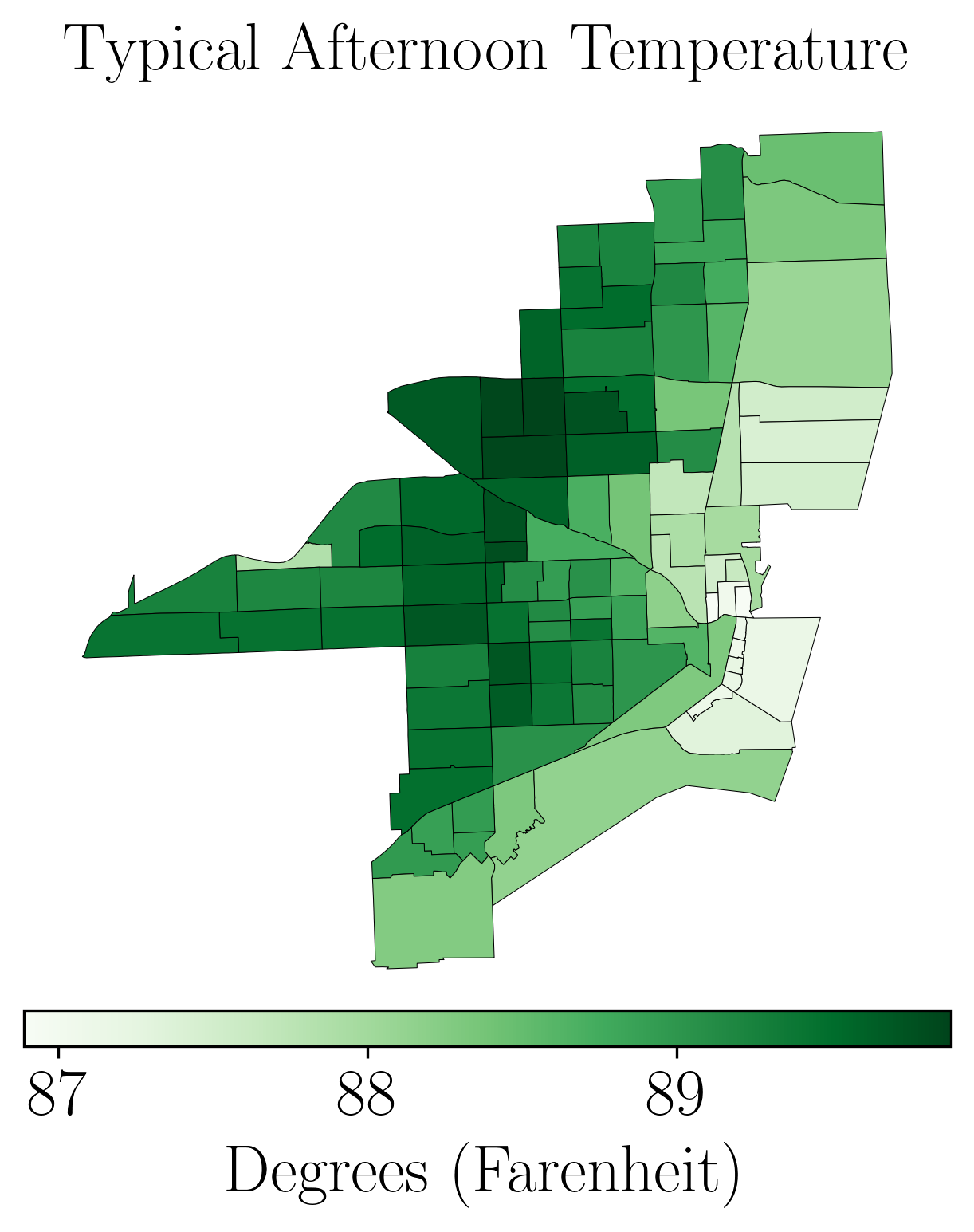} 
    \includegraphics[width=0.24\linewidth]{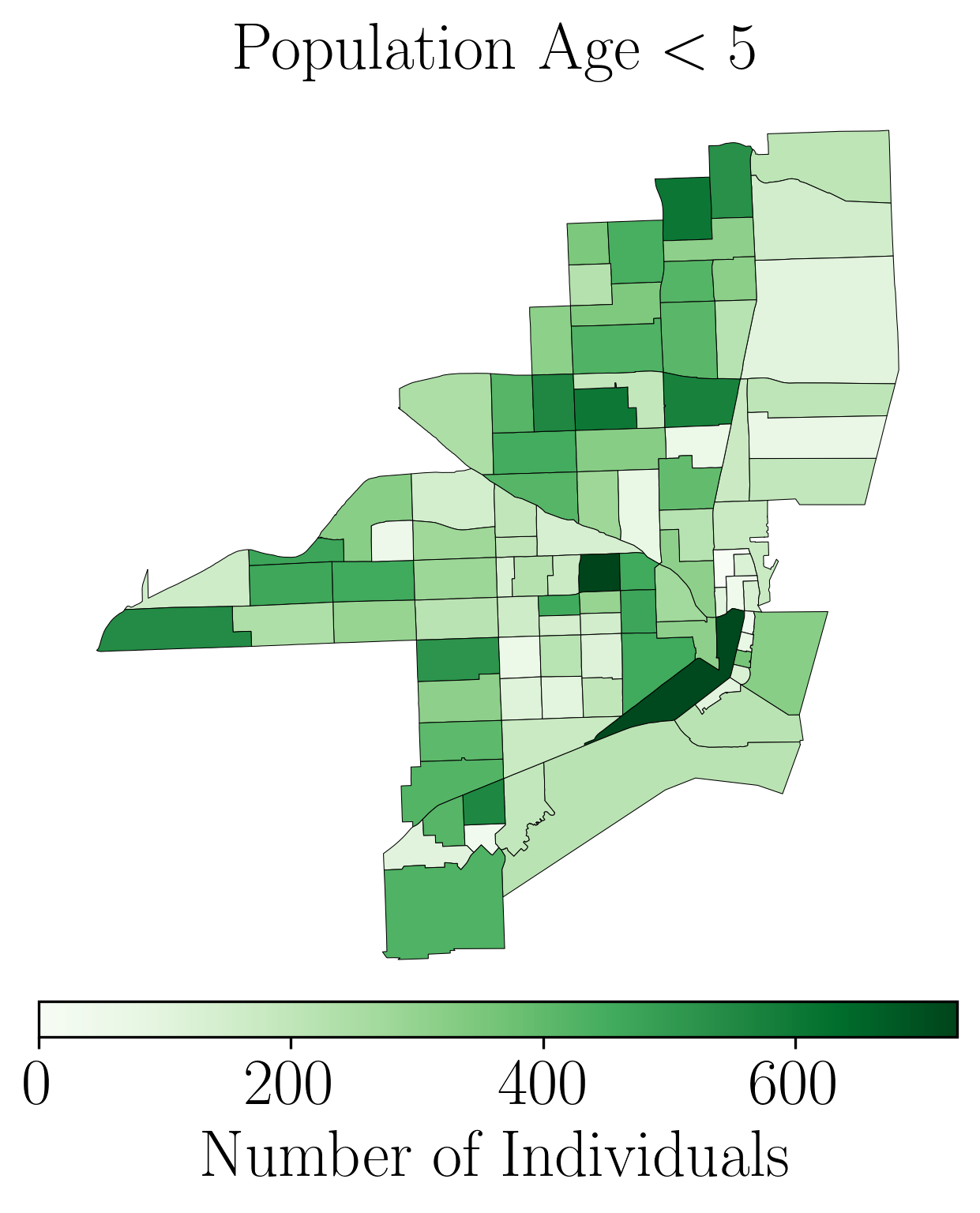} 
    \includegraphics[width=0.24\linewidth]{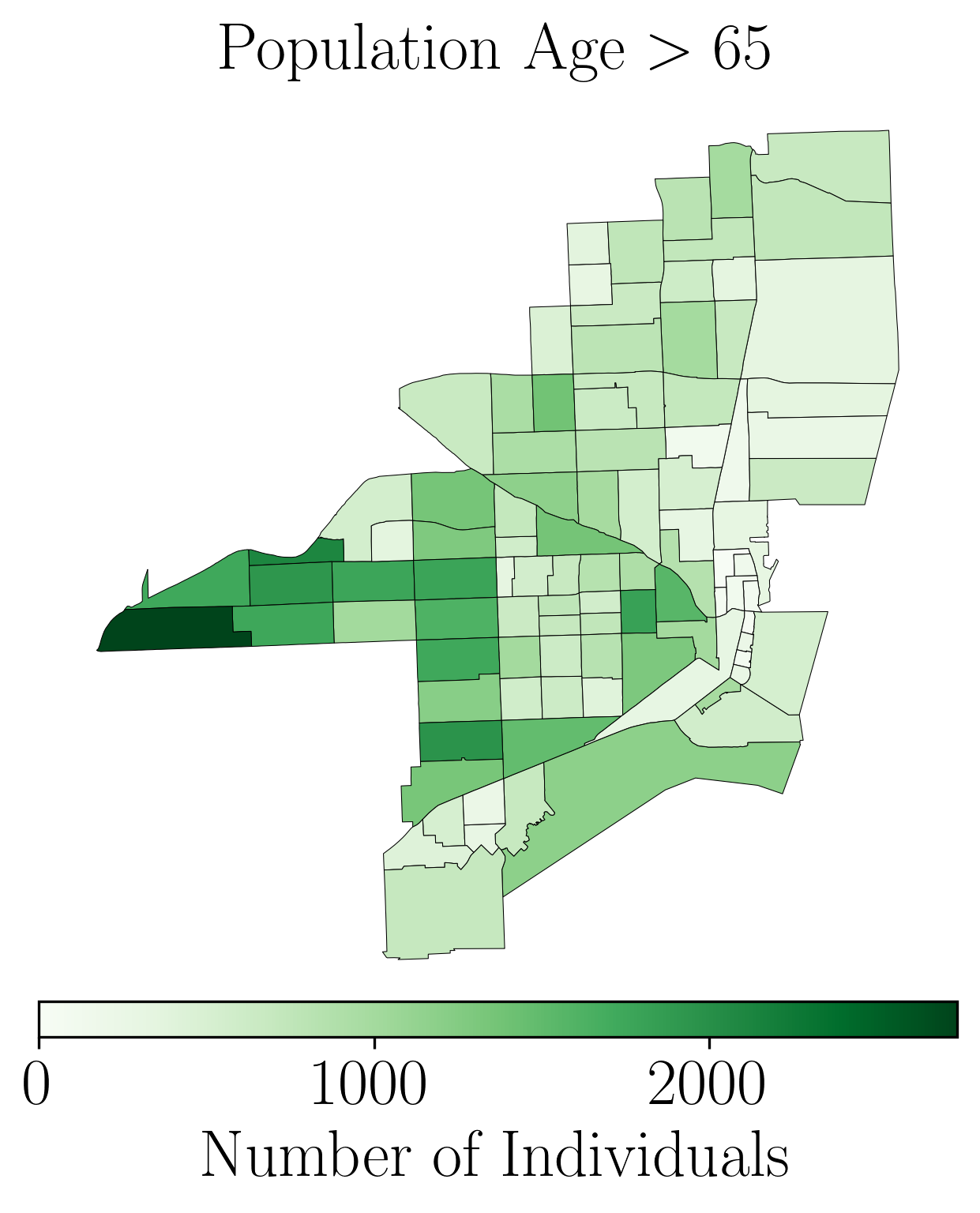}  \\

\vspace{0.3cm}

% \begin{figure}
    \centering
    {\large Portland} \\
    \includegraphics[width=0.24\linewidth]{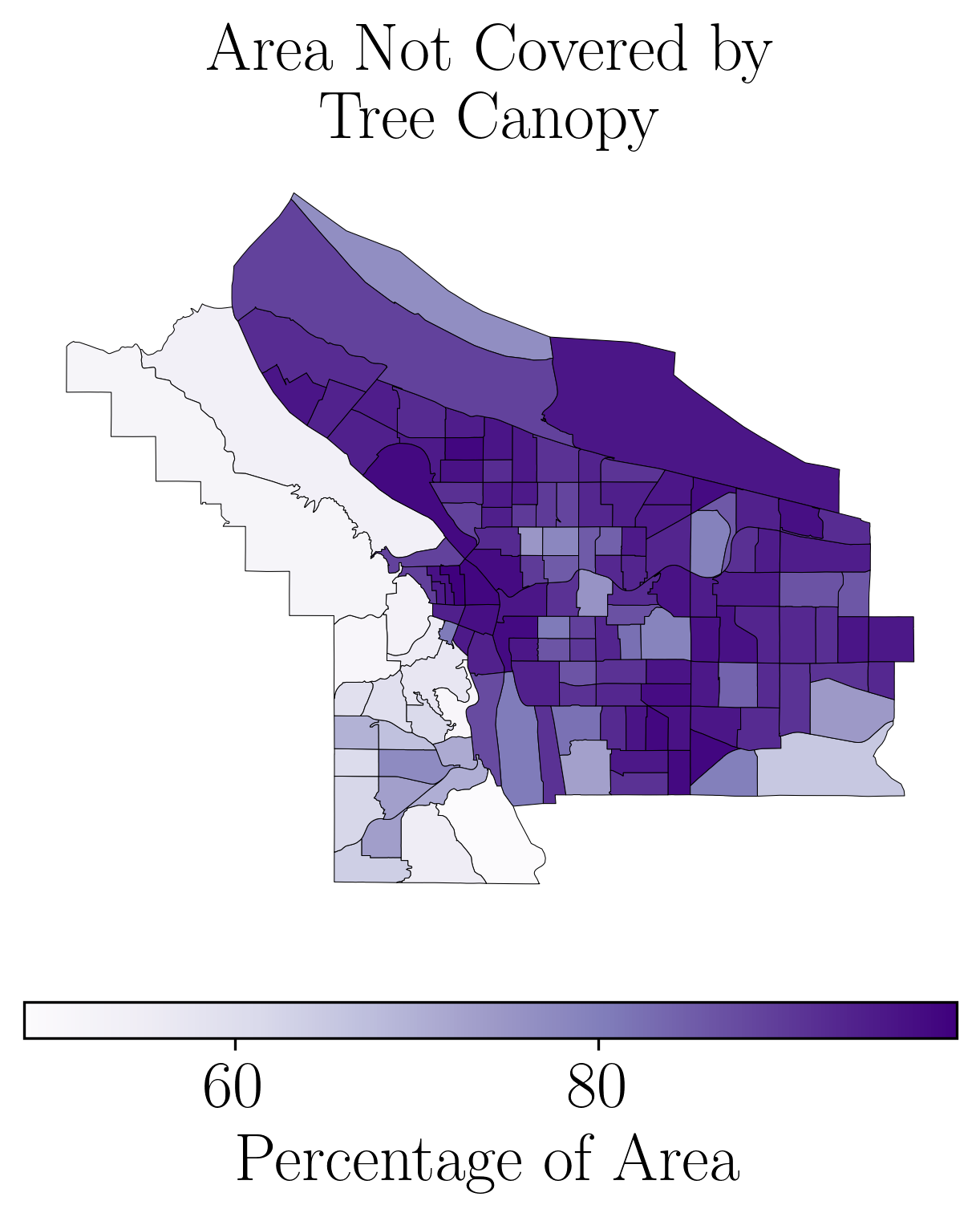}
    \includegraphics[width=0.24\linewidth]{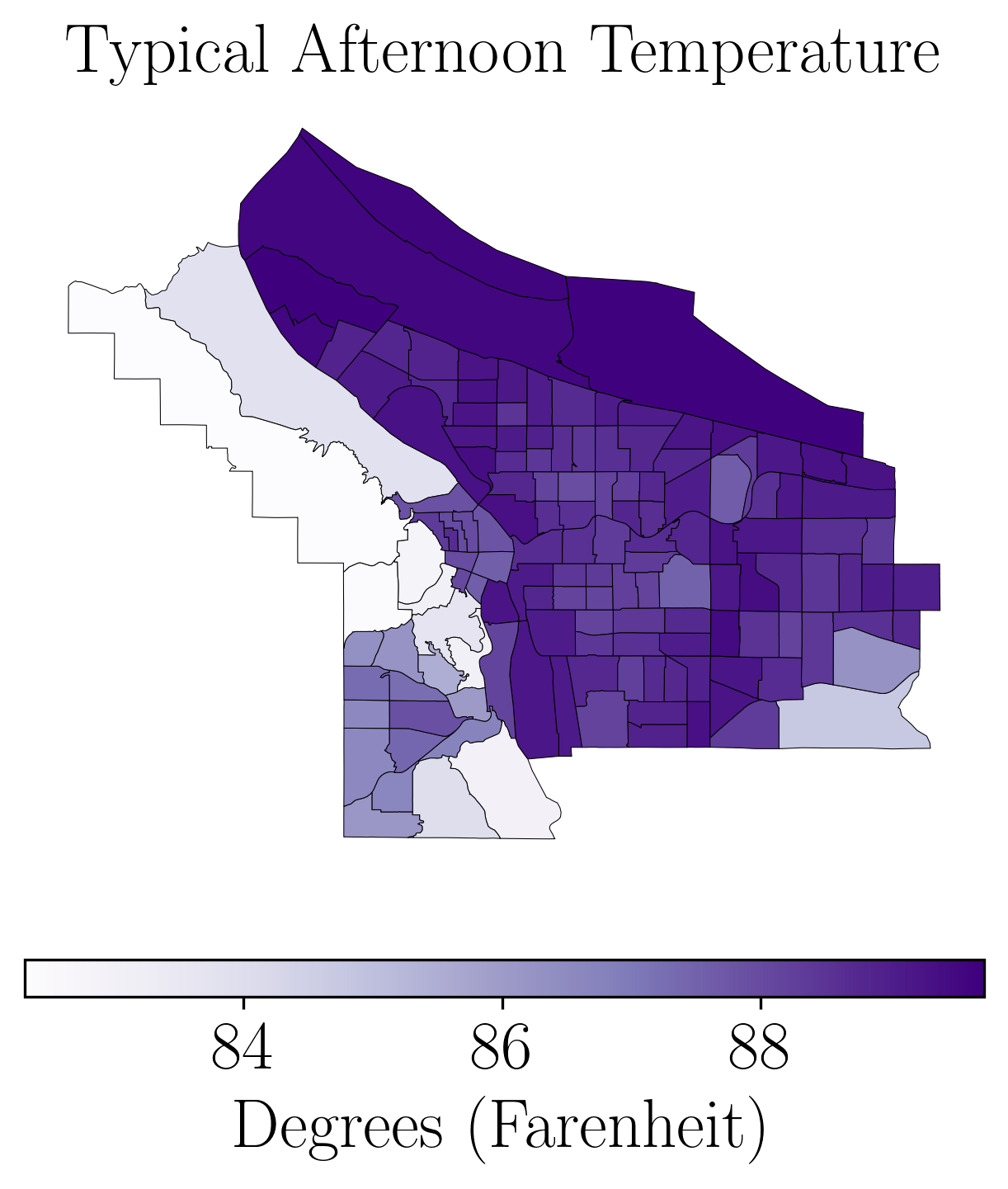}
    \includegraphics[width=0.24\linewidth]{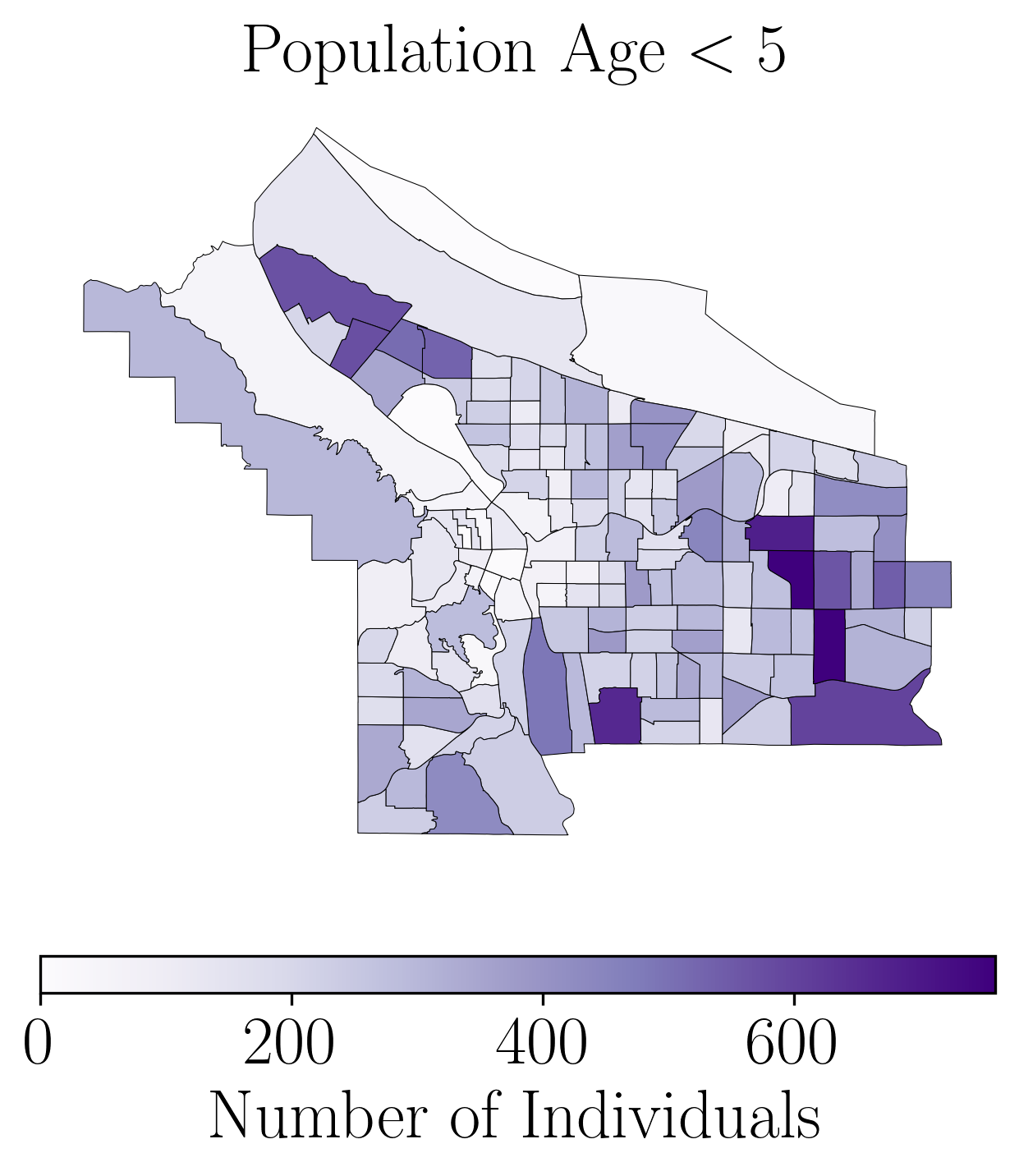} 
    \includegraphics[width=0.24\linewidth]{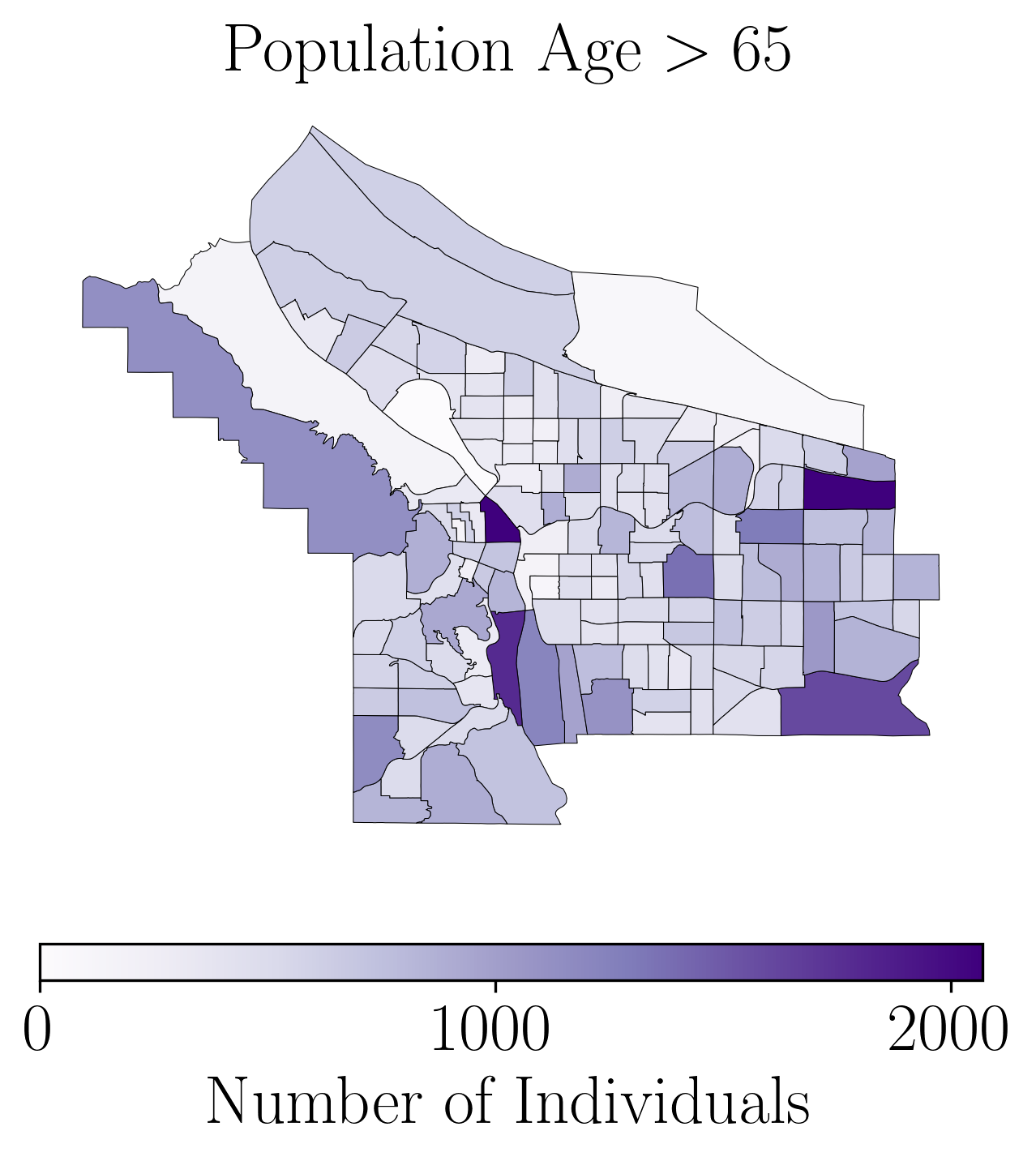} 
    \caption{Heatmaps of the four HVI variables for Austin, Boston, Miami, and Portland. From left to right: area not covered by tree canopy, typical afternoon temperature, number of individuals of age less than 5 years old, and number of individuals of age greater than 65.}
    \label{fig:HVI_figs_all}
\end{figure}

The following Shapefiles were downloaded in February 2024 to create the HVI maps:

\begin{enumerate}[itemsep=2pt,leftmargin=0.7cm]
\item TIGER/Line Shapefile, Current, State, Massachusetts, Census Blocks
\item TIGER/Line Shapefile, 2021, State, Texas, Census Blocks
\item TIGER/Line Shapefile, 2021, State, Oregon, Census Blocks 
\item TIGER/Line Shapefile, 2021, State, Florida, Census Blocks 
\end{enumerate}
Data on the typical afternoon temperature, the percentage of the area covered in tree canopy, number of individuals who are under 5, and number of individuals who are older than 65 was retrieved from the National Integrated Heat Health Information System (NIHHIS). The dataset was created as part of the National Oceanic and Atmospheric Administration's (NOAA's) ``U.S. Urban Heat Island Mapping Campaign." Data on the area \textit{not} covered in tree canopy was found by subtracting the data on the percent of area covered in tree canopy in a given census tract from 100. 
Figure~\ref{fig:HVI_figs_all} shows the individual HVI variables for each city.

Our motivation for incorporating these particular four variables into our measure of vulnerability is detailed as follows. 
Firstly, one can imagine that a warmer census block with few trees may not fare well during an extreme heat event. 
Secondly, being over the age of 65 puts an individual more at risk of heat related mortality due to social isolation and a lack of mobility to get assistance when faced with emergencies \cite{CVAReport}. 
Lastly, because children thermoregulate less effectively than adults, they are more likely to respond poorly during extreme heat events \cite{CVAReport}. 
The census block is interpreted to be at greater risk when its HVI is higher.

\section{Additional Data Analysis} \label{appendix:additional_data_analysis}
\begin{figure}
    \centering
    \includegraphics[width=0.45\linewidth]{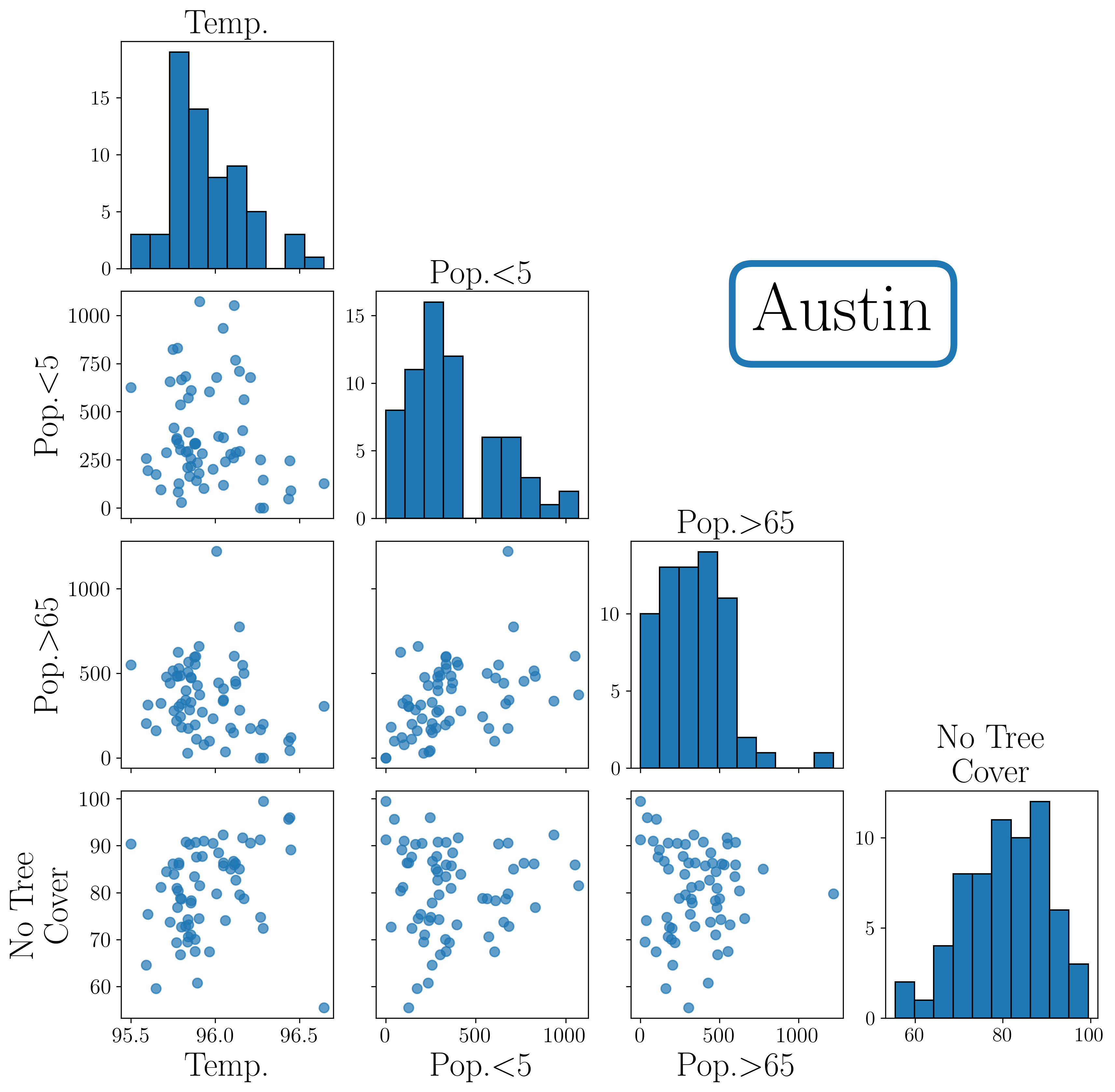}
    \includegraphics[width=0.45\linewidth]{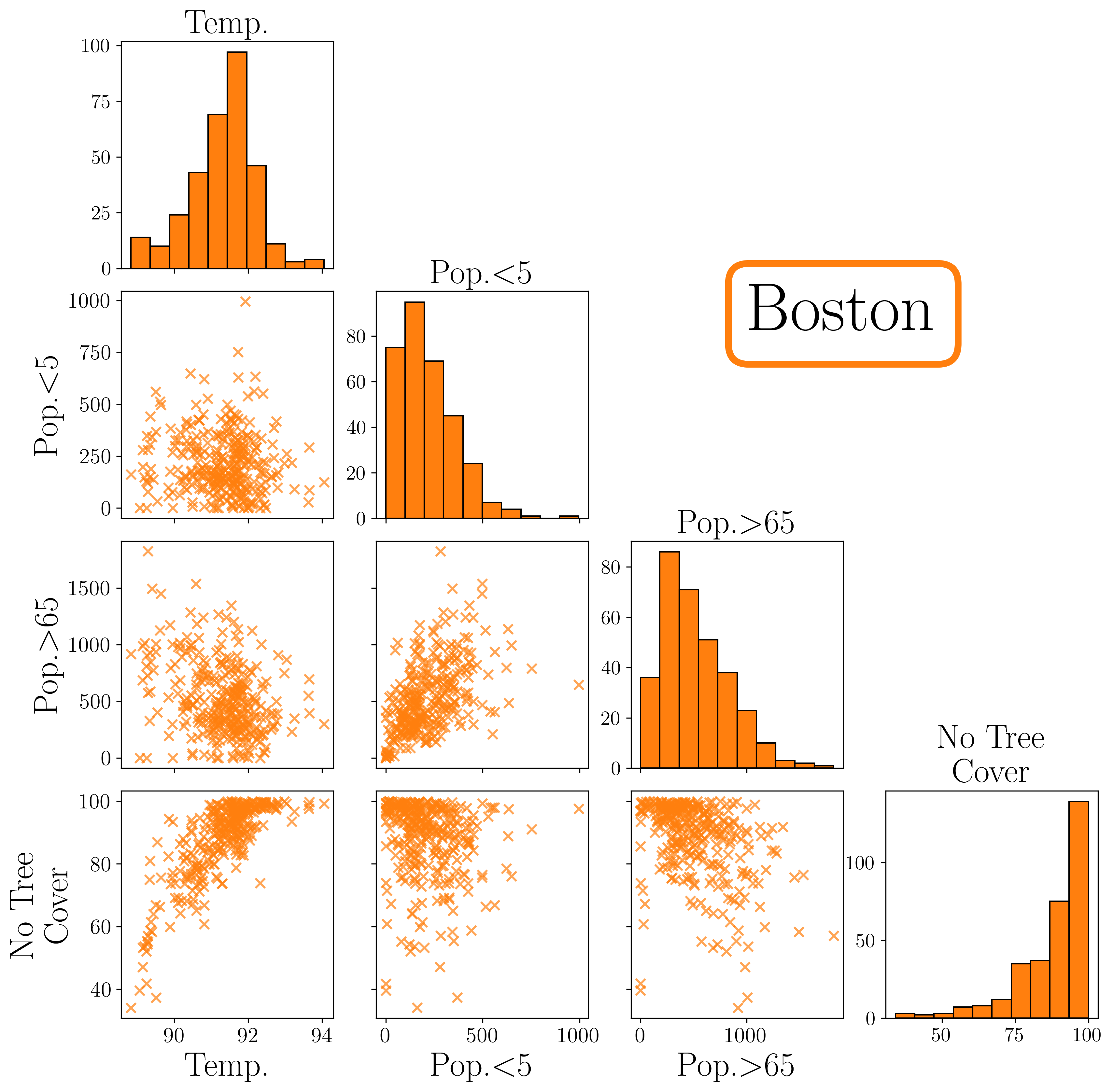} \\ 
    \includegraphics[width=0.45\linewidth]{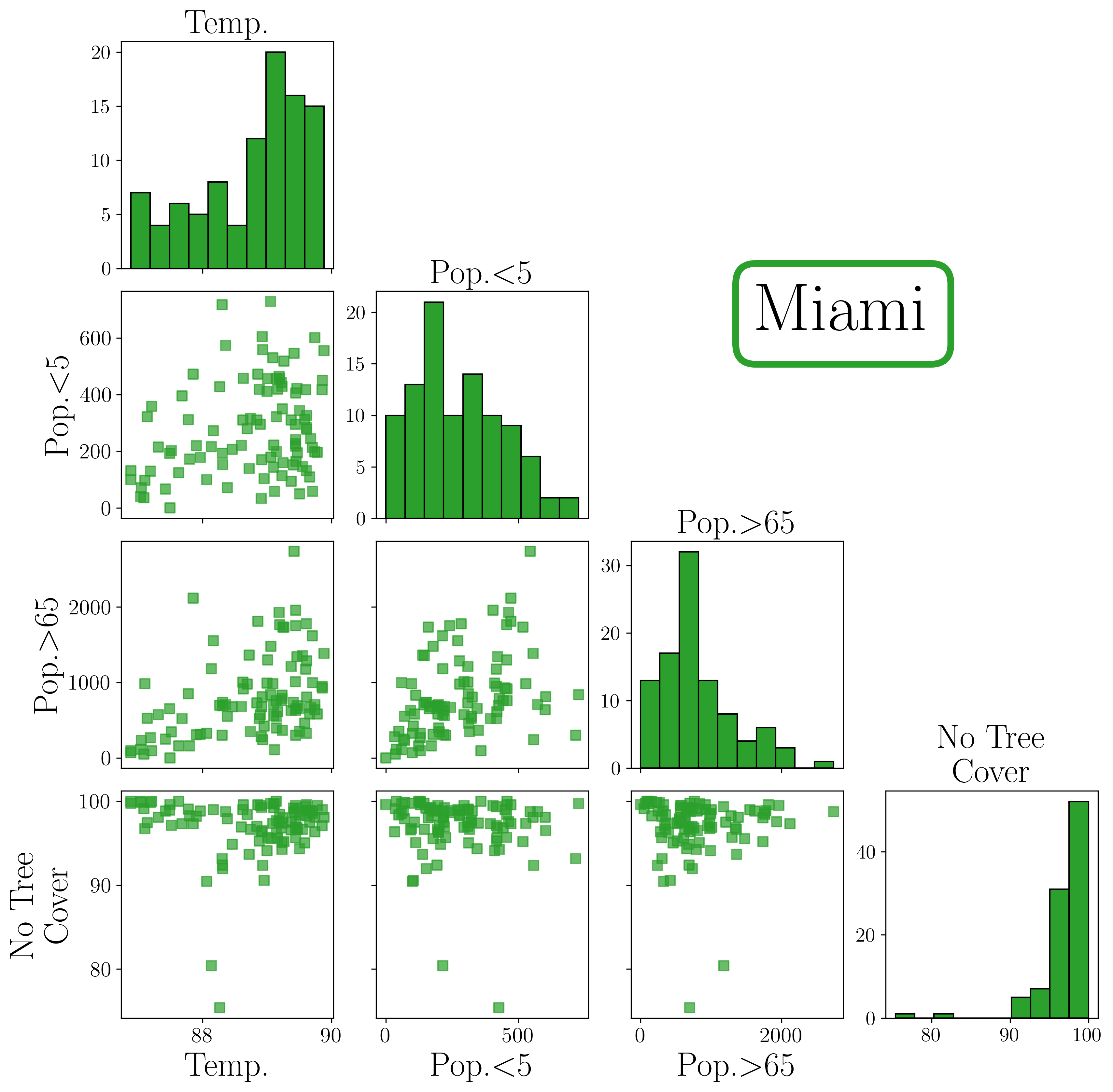}
    \includegraphics[width=0.45\linewidth]{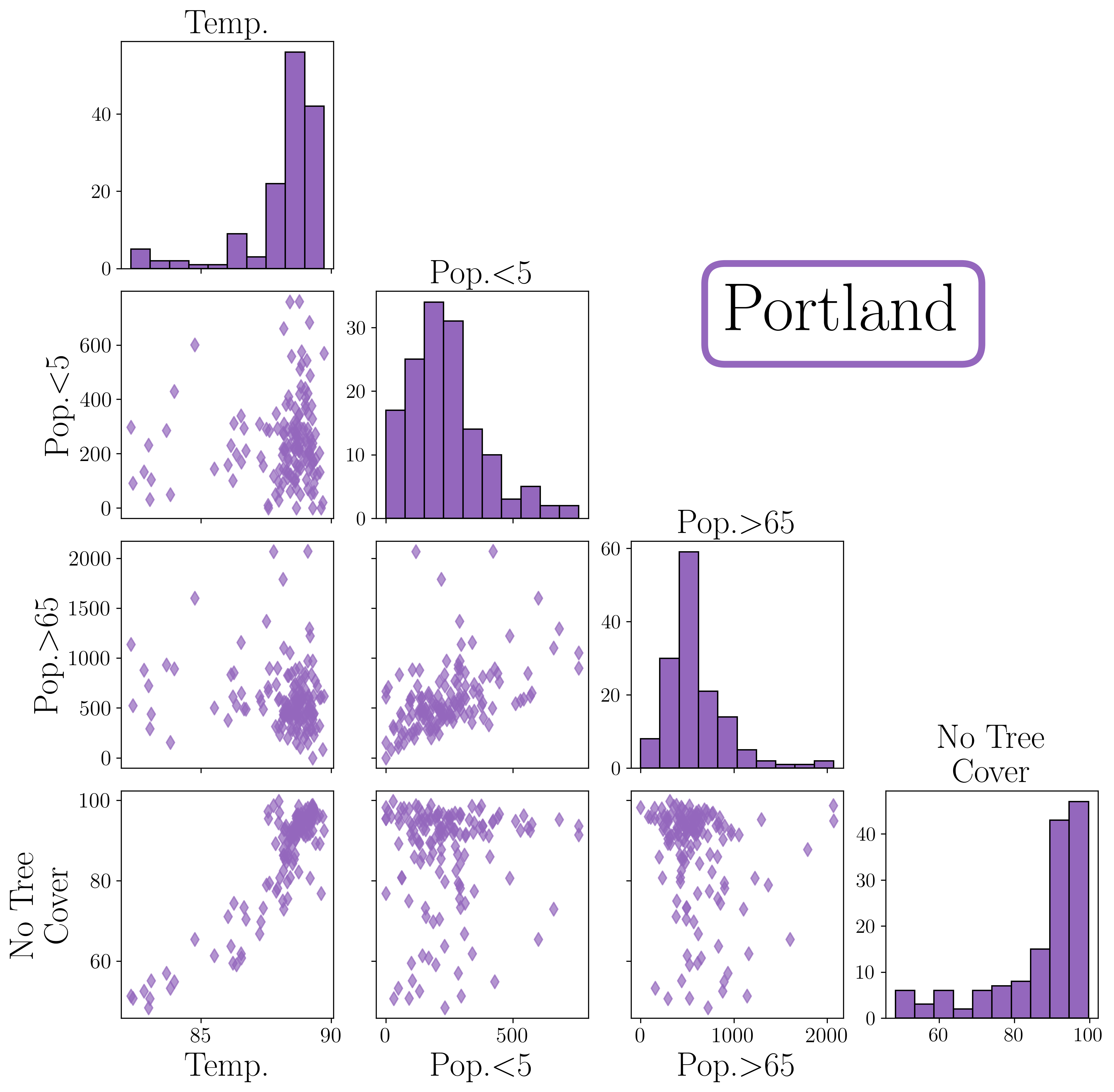}
    \caption{Histograms and pairwise scatter plots of HVI variables across census tracts within each city.}
    \label{fig:HVI_histograms}
\end{figure}

\begin{figure}
    \centering
    \includegraphics[width=\linewidth]{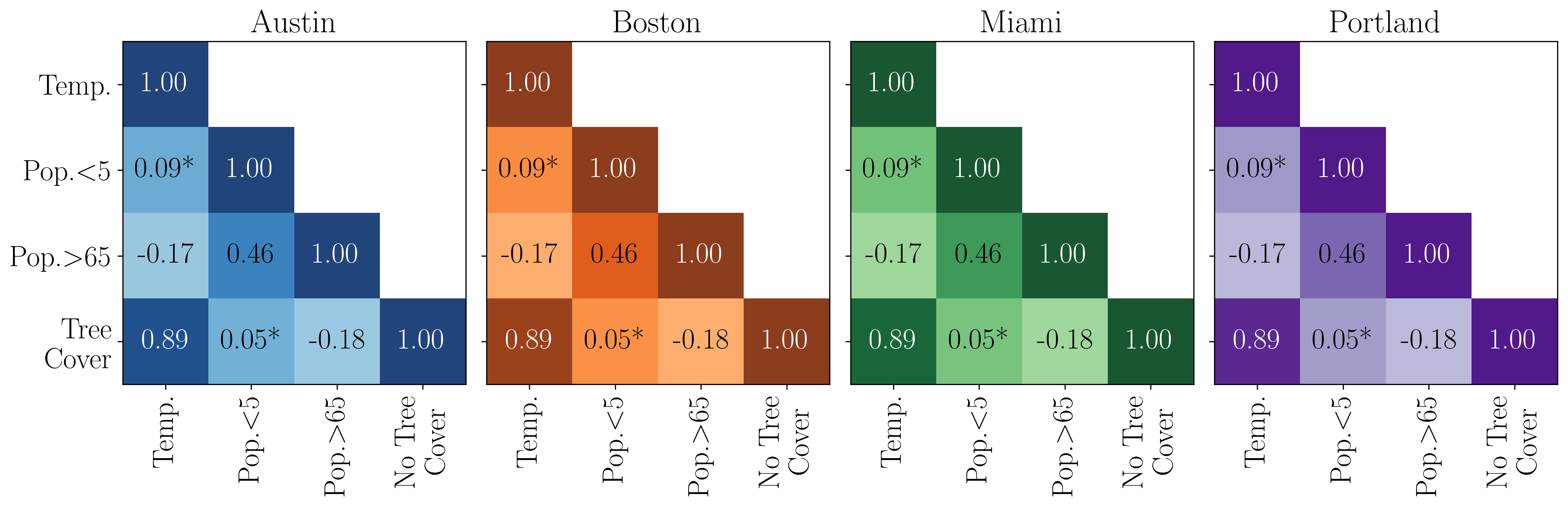}
    \caption{Matrices of pairwise Pearson correlations between the specified variables within each city. A star indicates the correlation is not statistically significant according to the two-sided $t$-test as it does not have a $p$-value  of $p< 0.05$.}
    \label{fig:HVI_correlationmatrices}
\end{figure}

We tested the independence of the four variables from the HVI using variance inflation factor (VIF). 
The results are presented in Table \ref{table:VIF}. 
These results are discussed in Section~\ref{sssec:multicol}.
We further explore the relationships between the HVI variables by considering their pairwise correlations.
Figures ~\ref{fig:HVI_histograms} and \ref{fig:HVI_correlationmatrices} show scatterplots of all pairwise relationships and the corresponding correlation coefficients.
As expected, from the VIF analysis, Boston and Portland have a strong pairwise correlation between the typical afternoon temperature and the percentage of the area not covered with tree canopy. 
In Portland, however, neither variable is strongly correlated with either of the demographic variables. 

\begin{table}[!ht]
\caption{Variance inflation factor (VIF) results to check multicollinearity in the four cities of interest. 
Note: a VIF of 1 indicates no multicollinearity. VIFs between 1 and 3 are considered to have negligible to minimal multicollinearity \cite{multivar_data_analysis}.}
\centering
\begin{tabular}{l c c c c}
\hline\hline
\textbf{Variable} & \textbf{Austin} & \textbf{Boston} & \textbf{Miami} & \textbf{Portland} \\ %[0.5ex] % inserts table %heading 

\hline
Typical Afternoon Temp. & 1.1452 & 2.5749 & 1.2388 & 4.9190\\
\% No Tree Canopy & 1.0786 & 2.6938 & 1.0098 & 4.8477\\
Population age $<$ 5 & 1.2158 & 1.4188 & 1.2230 & 1.3257\\
Population age $>$ 65 & 1.2612 & 1.6367 & 1.3503  & 1.3527 \\
\hline
\end{tabular}
\label{table:VIF}
\end{table}

\bibliographystyle{siamplain}
\bibliography{references}
\end{document}